\documentclass[APA,LATO1COL]{WileyNJD-v2}

\articletype{Article Type}%
\received{26 April 2016}
\revised{6 June 2016}
\accepted{6 June 2016}

\usepackage{amssymb}
\usepackage{hyperref}
\usepackage{subfig}

\usepackage{hyperref}       
\hypersetup{
    colorlinks=true,
    linkcolor=blue,
    filecolor=magenta,
    citecolor=blue,
    urlcolor=cyan  
}
\numberwithin{equation}{section}  

\newcommand{\bX}{\mathbf{X}}  
\newcommand{\bY}{\mathbf{Y}}
\newcommand{\bZ}{\mathbf{Z}}
\newcommand{\bA}{\mathbf{A}}
\newcommand{\bB}{\mathbf{B}}
\newcommand{\bU}{\mathbf{U}}
\newcommand{\bD}{\mathbf{D}}
\newcommand{\bV}{\mathbf{V}}
\newcommand{\bH}{\mathbf{H}}  
\newcommand{\bW}{\mathbf{W}}
\newcommand{\bK}{\mathbf{K}}
\newcommand{\bI}{\mathbf{I}}
\newcommand{\bM}{\mathbf{M}}
\newcommand{\bN}{\mathbf{N}}

\newcommand{\bx}{\mathbf{x}}
\newcommand{\by}{\mathbf{y}}
\newcommand{\bw}{\mathbf{w}}

\newcommand{\bu}{\mathbf{u}}

\newcommand{\bmalpha}{\boldsymbol{\alpha}}
\newcommand{\bmbeta}{\boldsymbol{\beta}}
\newcommand{\bmxi}{\boldsymbol{\xi}}

\DeclareMathOperator*{\argmin}{arg\,min}
\DeclareMathOperator*{\tr}{tr}

\DeclareMathOperator*{\sign}{sign}

\newcommand{\norms}[1]{\left\lVert#1\right\rVert}
\newcommand{\normsFrob}[1]{\left\lVert#1\right\rVert_{F}} 

\begin{document}

\title{Feature Grouping and Sparse Principal Component Analysis with Truncated Regularization 
}

\author[1]{Haiyan Jiang}

\author[2]{Shanshan Qin*}

\author[3]{Oscar Hernan Madrid Padilla}

\authormark{Jiang \textsc{et al}}

\address[1]{\orgdiv{School of Statistics and Data Science}, \orgname{Nankai University}, \orgaddress{\state{Tianjin}, \country{China}}. \texttt{Email: jianghaiyan.cn@gmail.com} }

\address[2]{\orgdiv{School of Statistics}, \orgname{ Tianjin University of Finance and Economics}, \orgaddress{\state{Tianjin}, \country{China}}. \texttt{Email: qinsslzu@gmail.com}}

\address[3]{\orgdiv{Department of Statistics}, \orgname{University of California, Los Angeles}, \orgaddress{\state{CA}, \country{United States}}. \\ \texttt{Email: oscar.madrid@stat.ucla.edu}}

\corres{*Shanshan Qin. \email{qinsslzu@gmail.com}}



\abstract[Summary]{
    In this paper, we consider a new variant for principal component analysis (PCA), {aiming to capture the grouping and/or sparse structures} of factor loadings simultaneously.
	To achieve these goals, we employ a non-convex truncated regularization with naturally adjustable sparsity and grouping effects, and propose the \underline{F}eature \underline{G}rouping and \underline{S}parse \underline{P}rincipal \underline{C}omponent \underline{A}nalysis (FGSPCA). 
	The proposed~\href{https://github.com/higeeks/FGSPCA}{FGSPCA}
	method encourages the factor loadings with similar values to collapse into disjoint homogeneous groups for feature grouping or into a special zero-valued group for feature selection, which in turn {helps reducing} model complexity and {increasing} model interpretation. 
	Usually, existing structured PCA methods require prior knowledge to construct the regularization term. However, the proposed FGSPCA can simultaneously capture the grouping and/or sparse structures of factor loadings without any prior information.
	To solve the resulting non-convex optimization problem, we propose an alternating algorithm that incorporates the difference-of-convex programming, augmented Lagrange method and coordinate descent method.	Experimental results demonstrate the promising performance and efficiency of the new method on both synthetic and real-world datasets.
	An R implementation of FGSPCA can be found on github~\href{ https://github.com/higeeks/FGSPCA}{https://github.com/higeeks/FGSPCA}.
	}
	

\maketitle

\section{Introduction}

Principal component analysis (PCA)~\citep{jolliffe1986principal} is an important unsupervised technique for feature extraction and dimension reduction, with numerous applications in statistics and machine learning, such as gene representation and face recognition.
The goal of PCA is to find a sequence of linear combinations of the original variables/predictors by projecting the original data onto an orthogonal linear space, called principal components (PCs), {such} that the derived PCs capture the maximum variance along the orthogonal direction.
Numerically, PCA can be obtained via the singular value decomposition (SVD) of the data matrix. 
Denote $\bX_{n\times p}\in \mathbb{R}^{n\times p}$ a data matrix consisting of $n$ observations of a random vector $\bx \in \mathbb{R}^{p}$ with a population covariance matrix $ \Sigma \in \mathbb{R}^{p\times p}$, where $n$ and $p$ are the number of observations and the number of variables/predictors, respectively.
Without loss of generality, assume that all the predictors are centered with $0$ means.
\normalem
Let the SVD of $\bX$ be $\bX=\bU\bD\bV^T$. The projection of the data $\bZ=\bU\bD (= \bX\bV)$ are the derived PCs, and the columns of $\bV$ are the corresponding \emph{factor loadings} (or \emph{factor coefficients}, or \emph{PC vectors}).

PCA aims to recover the top $k$ leading eigenvectors $\bu_1, \cdots, \bu_k$ of the population covariance matrix $\Sigma$, {with the corresponding eigenvalues $\lambda_1 \geq \cdots \geq \lambda_k$}. In high dimensional settings with $p \gg n$, the ordinary PCA can be inconsistent~\citep{paul2007asymptotics,nadler2008finite,johnstone2009consistency}, and additional assumptions are needed to avoid the  curse of dimensionality~\citep{wang2014tighten}.
Besides, a simple property of the ordinary PCA is that each PC usually involves all the original variables and the loadings (factor coefficients) are typically nonzero, which hinders the interpretability of the derived PCs.
In order to deal with the curse of dimensionality and improve the interpretability of the derived PCs, a sparsity assumption is often imposed on the loadings to get a sparsely weighted linear combination of the original variables. PCA with sparse loadings and its variants~\citep{vu2013fantope,zou2006sparse,cai2013sparse,erichson2020sparse} have been widely studied.
In the last decades, significant progress has been made on the methodological development as well as theoretical understanding of sparse PCA.
One can turn to~\citet{erichson2020sparse,wang2014tighten,jenatton2011structured,jenatton2010structured,grbovic2012sparse,croux2013robust,khan2015joint,yi2017joint,zou2018selective,jin2019principal,zhang2019robust,tian2019learning}, among others, for an overview of the literature.
Methods introduced in these articles intend to seek modified principal components for various sparsity properties. 
For example, SCoTLASS~\citep{jolliffe2003modified} is proposed by directly imposing an $\ell_1$ penalty on the ordinary PC vectors to get sparse loadings.
Sparse PCA (SPCA)~\citep{zou2006sparse} seeks sparse loadings by extending the elastic net~\citep{zou2005regularization} procedure and relaxing the orthogonality constraint of the ordinary PC vectors.

In addition to the sparsity property among loadings, structured grouping property can also lead to good interpretability of the resulting PCs.
SPCA~\citep{zou2006sparse} can achieve better interpretability by producing modified {PCs} with sparse loadings. However, it does not take into account the structured grouping property among loadings, i.e., clusters or groups. Based on the structured variable selection method~\citep{jenatton2011structured}, a structured sparse PCA~\citep{jenatton2010structured} is proposed to explore the structural information, as an extension of sparse PCA, and it incorporates prior knowledge into the sparsity-inducing regularization and is able to encode more sophisticated sparsity patterns.
In order to capture the `blocking' structures in the factor loadings, \citet{guo2010principal} proposed another variant of PCA with sparse fused loadings, named sparse fused PCA (SFPCA), by introducing a fusion penalty {that} encourages the loadings associated with high correlation to be close to get the `blocking' structures. 
Recently,~\citet{tian2019learning} proposed the feature-sparsity (row-sparsity) constrained PCA by considering feature-sparsity
structures for feature selection and PCA simultaneously.
However, these methods depend heavily on the structured prior knowledge which is usually challenging to obtain or specify in real applications. 
In~\citet{guo2010principal}, for example, the `blocking' structure is captured by the fusion penalty, where the fusion penalty depends on the sample correlation which serves as the prior information. Moreover, even though the PC vectors derived from the structured sparse PCA possess some sparse structures, they suffer from the same issue, that is, the structured sparsity depends on the given structural prior information.

In the ordinary PCA, each PC is a linear combination of all $p$ variables, and the loadings are typically nonzero and have no grouping effect. 
As is discussed above, the loadings can be sparse in sparse PCA~\citep{zou2006sparse}, but  dismissing grouping effect or clustering effect among the loadings.
In structured PCA~\citep{guo2010principal, jenatton2010structured,tian2019learning}, the structures of the loadings can be learned based on the structural prior knowledge/information which should be given to construct the regularization term in these methods.

In this paper, we propose a new variant of PCA, named feature grouping and sparse principal component analysis (FGSPCA), which can simultaneously capture the grouping and sparse structure of factor loadings, leading to modified PCs with grouping- and sparse-guided loadings. By adopting the fact that PCA can be formulated as a regression-type optimization problem, the grouping- and sparse-guided loadings are obtained by imposing the grouping and sparsity constraints on the regression coefficients. 
We make the following contributions.
\begin{itemize}
	\setlength{\topsep}{-0.5em} 
	\setlength{\partopsep}{-0.5em}
	\item To our knowledge, we initially consider simultaneously the grouping effect as well as the sparsity effect among factor loadings of PCA in the absence of prior knowledge. 
	The proposed FGSPCA method achieves the goal of feature grouping and feature selection through regularization, whose construction does not depend on any prior knowledge. The grouping and sparsity structure is learned naturally from the model rather than from given prior information. 
	\item 
	The proposed FGSPCA method imposes a non-convex regularization term with naturally adjustable sparsity and grouping effect. We solve the non-convex FGSPCA problem approximated by a sequence of linear convex subproblems {via the difference-of-convex programming (DC)}. Each of the convex subproblems is solved iteratively by incorporating {augmented Lagrange method (AL), and coordinate descent method (CD).} 
	\item The experiments on both synthetic and real-world data  demonstrate the promising performance of the proposed FGSPCA method.
\end{itemize}
Through out this paper, we use the following notations. 
Bold-face lower-case letters refer to vectors, e.g. $\bmalpha,\bmbeta$, and bold-face upper-case letters refer to matrices, e.g. $\bA, \bB$. 
{For a vector $\bw \in \mathbb{R}^{p}$, denote by  $\|\bw\|_2^2 = \sum_{j=1}^{p} w_j^2$, $\|\bw\|_1 = \sum_{j=1}^{p} |w_j|$, and $\|\bw\|_\infty = \max_{j \in \{1,\ldots,p\}}\{|w_j|\}$ the squared $\ell_2$ norm, the $\ell_1$ norm, and the maximum norm, respectively.}
For a matrix $\bW$, let $\|\bW\|_F = \sqrt{ \sum_{i,j} w_{ij}^2}$ denote the Frobenius norm. Note that ${ \|\bW\|_F^2 = \tr(\bW^T\bW) }$. 
Denote by $\bI_{k \times k}$ the identity matrix in $\mathbb{R}^{k \times k}$. Let $\mathrm{I}_{\{\mathcal{A}\}} = 1 $ if the condition $\mathcal{A}$ holds, otherwise $\mathrm{I}_{\{\mathcal{A}\}} =  0$.

The rest of the paper is organized as follows. In~\autoref{sec:SPCA}, the PCA is revisited.~\autoref{sec:FGSPCA} introduces the proposed FGSPCA and its connections to other sparse PCA variants. We propose an alternating algorithm to solve the FGSPCA problem in~\autoref{sec:FGS-algs}. Experiments to show the performance of FGSPCA and comparisons with other dimension reduction methods are presented in~\autoref{sec:simulation}. A discussion on the extension of FGSPCA to the settings with non-negative loadings falls into~\autoref{sec:discussion}. We conclude the paper in~\autoref{sec:conclusion}.

\section{Principal Component Analysis Revisited}\label{sec:SPCA}
Let $\bX=(x_{ij})_{n \times p}$ denote a data matrix with $n$ observations and $p$ variables. Assume that the columns of $\bX$ are all centered.
In PCA, each PC is obtained by constructing linear combinations of the original variables that maximize the variance. Denote the SVD of $\bX$ by $\bX=\bU \bD\bV^T$. Let $\bZ_j=\bU_j \bD_{jj}$ be the $j$-th PC, and columns of $\bV=[\bV_1, \cdots,\bV_k]$ be the PC vectors or PC loadings. Except for the SVD decomposition, another way to derive the PC vectors is to solve the following constrained least squares problem,
\begin{equation}\label{eq:ls-pca}
\min_{\bA } 
\|\bX - \bX\bA\bA^T \|_2^2 \ , \text{ s.t. } \bA^T \bA = \bI_{k\times k} \ ,
\end{equation}
where $\bA=[\bmalpha_1, \cdots, \bmalpha_k]$ is a $p \times k$ matrix with orthogonal columns.
The estimated $\widehat{\bA}$ contains the first $k$ PC vectors, and the projection of the data $\widehat{\bZ} = \bX \widehat{\bA}$ are the first $k$ PCs.

By relaxing the orthogonality requirement and imposing an $\ell_2$ penalty, 
~\cite{zou2006sparse} proposed and reformulated PCA as the following regularized regression optimization problem which is defined in~\autoref{lemma:one-pca} (Theorem 3 in~\citet{zou2006sparse}, see {\color{blue}{Appendix A}} in Supporting information).
\begin{lemma}\label{lemma:one-pca}
	Consider the first $k$ principal components. Let $\bx_i$ be the $i$-th row of data matrix $\bX$. Denote $\bA_{p\times k} = [\bmalpha_1, \cdots, \bmalpha_k]$, $\bB_{p\times k} = [\bmbeta_1, \cdots, \bmbeta_k]$. 
	For any $\lambda > 0$, let
	\begin{equation}\label{eq:one-pca}
	(\widehat{\bA}, \widehat{\bB}) = \argmin_{\bA, \bB} 
	\sum_{i=1}^{n} \|\bx_i - \bA\bB^T \bx_i\|_2^2 + \lambda\sum_{j=1}^{k}\|\bmbeta_{j}\|_2^2\ , \text{ s.t. } \bA^T \bA = \bI_{k\times k} \ .
	\end{equation}
	Then $\widehat{\bmbeta}_{j} \propto \bV_{j}$\quad for $j=1, \cdots, k$.
\end{lemma}
The PCA problem is transformed into a regression-type optimization problem with orthonormal constraints on $\bA$, and all the sequences of principal components can be derived through~\autoref{lemma:one-pca}.
With the restriction $\bB=\bA$, the objective function becomes
$ \sum_{i=1}^{n} \|\bx_i - \bA\bB^T\bx_i\|_2^2
= \sum_{i=1}^{n} \|\bx_i - \bA\bA^T\bx_i\|_2^2 $,
whose minimizer under the orthonormal constraint on $\bA$ consists exactly of the first $k$ PC vectors of the ordinary PCA. Lemma
\autoref{lemma:one-pca} shows that the exact PCA can still be obtained by relaxing the restriction $\bB=\bA$ and adding the ridge penalty term.

Note that
\begin{equation*}
    \sum_{i=1}^{n} \norms{\bx_i - \bA\bB^T\bx_i}_2^2
    = \norms{\bX - \bX\bB\bA^T}_F^2 .
\end{equation*}
Since $\bA$ is orthonormal, let $\bA_{\perp}$ be any orthonormal matrix such that $[\bA; \bA_{\perp}]$ is $p\times p$ orthonormal. 
Then we have
$ { \normsFrob{\bX - \bX\bB\bA^T}^2 
	=  \sum_{j=1}^{k} \|\bX \bmalpha_j - \bX\bmbeta_j\|_2^2 +  \normsFrob{\bX\bA_{\perp} }^2 } $.
Suppose that $\bA$ is given, then the optimal $\bB$ can  be obtained by minimizing 
$ { \sum_{j=1}^{k} \|\bX \bmalpha_j - \bX\bmbeta_j\|_2^2 + \lambda \sum_{j=1}^{k} \|\bmbeta_j\|_2^2 } $, 
which is equivalent to $k$ independent ridge regression problems. 

\section{The Methods}\label{sec:FGSPCA}

\subsection{Feature Grouping and Sparse Loadings}
\begin{figure}[http]
	\centering
	\subfloat[][\centering Penalty Functions ] {\includegraphics[width=0.5\textwidth]{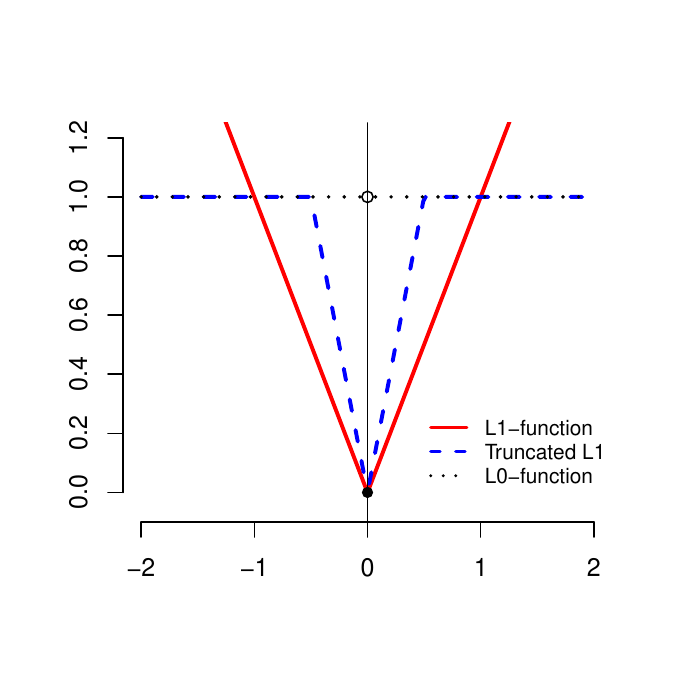}} %
	\subfloat[][\centering Thresholding Functions ] {\includegraphics[width=0.5\textwidth]{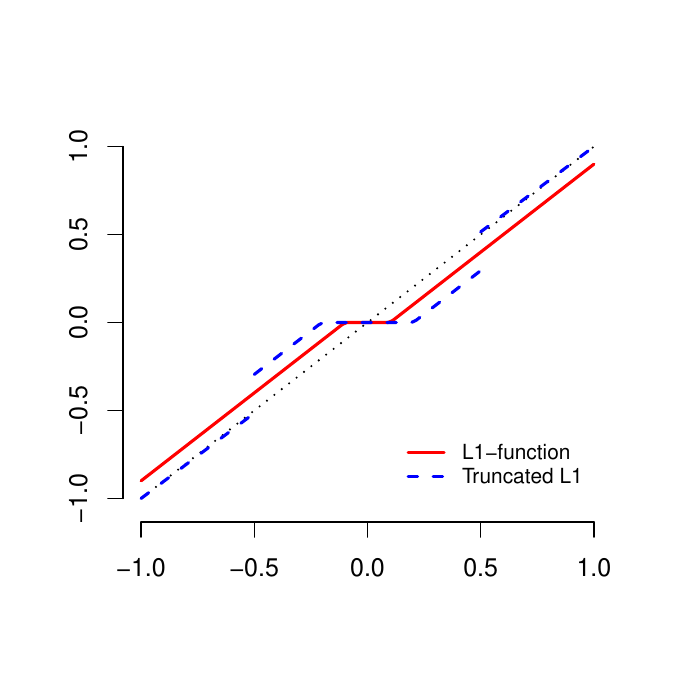}}%
	\caption{Comparison of different penalty functions (left panel): the $\ell_1$-function (solid line), the truncated $\ell_1$-function (dashed line), and the $\ell_0$-function (dotted line), and their corresponding thresholding functions (right panel). The truncated $\ell_1$-function $\min\{\frac{|x|}{\tau}, 1\}$ approximates the $\ell_0$-function $I(x \neq 0)$ as $\tau \rightarrow 0$, and it is closer to the $\ell_0$ penalty than the $\ell_1$ penalty. The thresholding functions show that, compared to the $\ell_1$ penalty, the truncated $\ell_1$ penalty penalizes more aggressively with small coefficients preferred, and it has no bias with large coefficients.}
	\label{fig:penalty}
\end{figure}

In order to investigate the structures among loadings, we extend the optimization problem~\eqref{eq:one-pca} by imposing feature grouping and feature selection penalties simultaneously, to get feature grouping and sparse loadings simultaneously.
The proposed \textbf{FGSPCA model} is based on solving the following optimization problem,
\begin{equation}\label{eq:FGSPCA-crit}
\min_{\bA, \bB} 
\sum_{i=1}^{n} \|\bx_i - \bA\bB^T \bx_i\|_2^2 
+ \lambda \sum_{j=1}^{k} \|\bmbeta_j\|_2^2  
+ \lambda_{1} \sum_{j=1}^{k} p_1(\bmbeta_j) 
+ \lambda_2 \sum_{j=1}^{k} p_2(\bmbeta_j)\ , \text{ s.t. } \bA^T \bA = \bI_{k\times k}, 
\end{equation}
where ${p}_{1}(\bmbeta)$ and ${p}_{2}(\bmbeta)$ are regularization functions, taking the following penalty forms, 
\begin{equation}\label{eq:truncated-pen}
p_1(\bmbeta_j) = \sum_{l=1}^{p} \min\left\{\frac{|\beta_{l(j)}|}{\tau}, 1\right\}, \quad
p_2(\bmbeta_j) = \sum_{l<l': (l, l') \in \mathcal{E}} \min\left\{ \frac{|\beta_{l(j)} - \beta_{l'(j)}|}{\tau}, 1\right\} \ ,
\end{equation}
where $\beta_{l(j)}$ denotes the $l$-th element of the vector $\bmbeta_{j}$.  $\lambda_1 (>0), \lambda_2 (>0)$ are the corresponding tuning parameters, and $\tau > 0$ is a thresholding parameter which determines when a small coefficient or a small difference between two coefficients will be penalized. The notation $\mathcal{E} \subset \{1, \cdots, p\}^2$ refers to  a set of edges on a fully connected and undirected graph (complete graph), with $l \sim l'$ indicating an edge directly connecting two distinct nodes $l \neq l'$, where each node represents a variable. 
\autoref{fig:penalty} gives a comparison of different penalty functions and their thresholding functions.
We refer the reader to  {\color{blue}{Appendix B}}  in Supporting information for more structured sparsity regularization functions.
\begin{remark}
	A). The key point of the FGSPCA with $p_1(\cdot)$ and $p_2(\cdot)$ penalty functions can be viewed as performing feature selection and feature grouping simultaneously.
	B). As shown in~\citet{shen2012simultaneous}, the truncated $\ell_1$-function $\min(\frac{|\beta_l|}{\tau}, 1)$ can be regarded as a non-convex and non-smooth surrogate of $\ell_0$-function $I(\beta_l\neq 0)$ when $\tau \rightarrow 0$. Besides, the selection consistency can be achieved by the  $\ell_0$-penalty and its surrogate---the truncated $\ell_1$-penalty~\citep{shen2013constrained,dai2021truncated}. Therefore, the sparse PCA with $\ell_1$ penalty cannot  achieve selection consistency. The intuition is that compared to the $\ell_1$ penalty, the truncated $\ell_1$ penalty is closer to the $\ell_0$ penalty and penalizes more aggressively with small coefficients preferred.
	Meanwhile, the truncated $\ell_1$-function $\min(\frac{|\beta_l|}{\tau}, 1)$ can be a good approximation of $\ell_1$-function as $\tau \rightarrow \infty$. 
	C). One may use the $\ell_1$-function $|\beta_l|$ as a smooth approximation of $\ell_0$-function. However, the shrinkage bias tends to be larger as parameter size gets larger~\citep{wu2018general,yun2019trimming} since the $\ell_1$ penalty is proportional to the size of parameters.
	The smooth approximation, $\ell_1$-function, has the drawback of producing biased estimates for large coefficients and lacking oracle property~\citep{fan2001variable,zhang2008sparsity}. 
\end{remark}
\subsection{Connection to Sparse PCA Variants}
By relaxing the orthogonality requirement and extending the elastic net procedure, the sparse PCA (SPCA)~\citep{zou2006sparse}  solves the following regularized optimization problem,
\begin{equation}\label{eq:sparse-pca}
\begin{split}
(\widehat{\bA}, \widehat{\bB}) = & \argmin_{\bA, \bB} 
\sum_{i=1}^{n} \|\bx_i - \bA\bB^T \bx_i\|_2^2 
+ \lambda \sum_{j=1}^{k}\|\bmbeta_{j}\|_2^2 
+ \lambda_{1} \sum_{j=1}^{k} \|\bmbeta_{j}\|_{1} \\ 
& \text{ subject to } \bA^T \bA = \bI_{k\times k}.
\end{split}
\end{equation}
Note that the optimization problem in~\eqref{eq:sparse-pca} is a special case of~\eqref{eq:FGSPCA-crit} as $\tau \to \infty$ and  $\lambda_2=0$. By imposing a fusion penalty, the sparse fused PCA (SFPCA) with sparse fused loadings~\citep{guo2010principal} solves the following regularized optimization problem, 
\begin{align}\label{eq:sparse-fused-pca}
\min_{\bA, \bB} 
& \sum_{i=1}^{n} \|\bx_i - \bA\bB^T \bx_i\|_2^2 
+ \lambda \sum_{j=1}^{k}\|\bmbeta_{j}\|_2^2 
+ \lambda_1 \sum_{j=1}^{k} \|\bmbeta_{j}\|_1
+ \lambda_2 \sum_{j=1}^{k} \sum_{s\neq t} |\rho_{s,t}| |\beta_{s(j)} - \sign(\rho_{s,t}) \beta_{t(j)}| \nonumber \\
& \text{ subject to } \bA^T \bA = \bI_{k\times k} \ , 
\end{align}
where $\rho_{s,t}$ denotes the sample correlation between variables $X_s$ and $X_t$, and $\sign(x)$ returns the sign of $x$.
For fair comparison, we add an $\ell_2$ penalty to the objective function of the SFPCA criterion.
The SFPCA~\citep{guo2010principal} can obtain sparse fused loadings in a more interpretive way, where the fusion penalty depends on the sample correlation,  serving as prior knowledge. Therefore, the SFPCA encourages the loadings associated with high correlation to have the same magnitude. 

\section{The Algorithms}\label{sec:FGS-algs}
\subsection{Alternating Optimization Algorithm of FGSPCA}
In this section, we discuss the algorithms to optimize the proposed objective function in \eqref{eq:FGSPCA-crit}. An alternating optimization algorithm over $\bA$ and $\bB$ is employed, analogously to the SPCA algorithm~\citep{zou2006sparse} and SFPCA algorithm~\citep{guo2010principal}.
Specially, the alternating algorithm to solve the optimization problem~\eqref{eq:FGSPCA-crit} proceeds as follows.\\
\textbf{Algorithm 1. The FGSPCA Algorithm.}
\begin{enumerate}
	\item[\emph{Step 1.}]
	Initialize $\widehat{\bA} $ by setting it to be $\bV[,1:k]$, the first $k$ ordinary PC vectors.
	\item[\emph{Step 2.}]
	(Estimation of $\bB$ given $\bA$).
	Given a fixed $\bA = [\bmalpha_1, \cdots, \bmalpha_k]$, minimizing the objective function~\eqref{eq:FGSPCA-crit} over $\bB$ is equivalent to solving the following $k$ separate subproblems, for $j=1,\cdots, k$,
	\begin{equation}\label{eq:givenA-solveB}
	\widehat{\bmbeta}_{j} = \argmin_{\bmbeta_j} \| \bZ_j - \bX \bmbeta_j\|_2^2
	+ \lambda \|\bmbeta_{j}\|_2^2
	+ \lambda_1 \sum_{l=1}^{p}\min \left(\frac{|\beta_{l(j)}|}{\tau}, 1\right) 
	+ \lambda_2 \sum_{l < l':(l,l') \in \mathcal{E}} \min \left(\frac{|\beta_{l(j)} - \beta_{l'(j)}|}{\tau}, 1\right) ,
	\end{equation}
	where $\bZ_j = \bX \bmalpha_j$.
	The optimization of~\eqref{eq:givenA-solveB} is discussed in~\autoref{sec:FGS-solveB}. 
	In this step, we update $\bB$ and obtain the estimate $\widehat{\bB}=[\widehat{\bmbeta}_{1}, \cdots, \widehat{\bmbeta}_{k}]$. 
	\item[\emph{Step 3.}]
	(Estimation of $\bA$ given $\bB$).
	Given a fixed $\bB = [\bmbeta_1, \cdots, \bmbeta_k]$, minimizing the objective function~\eqref{eq:FGSPCA-crit} over $\bA$ is equivalent to solving the following problem, 
	\begin{equation}\label{eq:givenB-solveA}
	\min_{\bA} \normsFrob{\bX - \bX\bB\bA^T}^2 
	\quad \mathrm{s.t. } \quad \bA^T\bA = \mathbf{I}_{k\times k}.
	\end{equation}
	The solution to~\eqref{eq:givenB-solveA} can be obtained through a reduced rank Procrustes Rotation (Theorem 4 in~\citet{zou2006sparse}, see {\color{blue} Appendix C} in Supporting information). We compute the SVD of $\bX^T\bX\bB$ as $\bX^T\bX\bB = \bU\bD\bV^T$, then the solution of~\eqref{eq:givenB-solveA} is derived by $\widehat{\bA} = \bU\bV^T$. 
	In this step, we update $\bA$ and obtain the estimate $\widehat{\bA}=[\widehat{\bmalpha}_{1}, \cdots, \widehat{\bmalpha}_{k}]$.
	\item[\emph{Step 4.}] Repeat \emph{Steps 2---3} until convergence.
\end{enumerate}

\begin{remark}
	A). The initialization of $\bA$, $\bV[, 1:k]$, can be loadings of any PCA method. For simplicity, let $\bV[, 1:k]$ be the first $k$ ordinary PC loadings. Clearly, $\bV[, 1:k]$ can also be initialized as the first $k$ PC loadings of  SPCA~\citep{zou2006sparse}, or the first $k$ PC loadings of SFPCA~\citep{guo2010principal}. 
	B). The convergence criterion in \emph{Step 4} can be verified by that the difference between two adjacent iterations of $\bB$ is small. {We use Frobenius norm to measure the matrix difference, that is, $ \|\bB_1 -\bB_2 \|_F^2 \leq \epsilon$, where $\epsilon$ is a small positive value, say, 1e-5.} 
\end{remark}

\subsection{Estimation of \texorpdfstring{$\bB$}{TEXT} given \texorpdfstring{$\bA$}{TEXT}}\label{sec:FGS-solveB}

Efficiently solving the subproblem~\eqref{eq:givenA-solveB} plays a key role in solving the problem~\eqref{eq:FGSPCA-crit}. 
The objective function~\eqref{eq:givenA-solveB} is a special case of a regularized regression problem with feature grouping and sparsity constraints (FGS). 
Thus this section gives an algorithm for the FGS problem, which is a core part of the Algorithm 1. 
The general form of the FGS problem is stated as follows, 
\begin{equation}\label{eq:LS-FGFS-general}
\min_{\bmbeta}
\sum_{i=1}^{n} (y_i - \bx_i^T \bmbeta)^2
+ \lambda \sum_{l=1}^{p}\beta_l^2
+ \lambda_1 \sum_{l=1}^{p}\min \left(\frac{|\beta_l|}{\tau}, 1\right) 
+ \lambda_2 \sum_{l < l':(l,l') \in \mathcal{E}} \min \left(\frac{|\beta_{l} - \beta_{l'}|}{\tau}, 1\right) .
\end{equation}
Since the above problem~\eqref{eq:LS-FGFS-general} is a non-convex optimization problem, we employ the  difference-of-convex programming (DC)~\citep{an2005dc}.
Our algorithmic solution for~\eqref{eq:LS-FGFS-general} is an extension of the algorithms in~\citet{shen2012simultaneous,qin2020high} by adding the $\ell_2$ penalty.
Our main technical contribution is to extend the algorithm in~\citet{shen2012simultaneous,qin2020high} to applications of developing more interpretable PCA.

We propose an integrated algorithm for the estimation of $\bB$ given $\bA$ (algorithm 1) which integrates the difference-of-convex algorithm (DC), the augmented Lagrange method and coordinate descent method (AL-CD), for efficient computation. The procedure to solve the FGS problem consists of three steps. \emph{First}, the non-convex objective function is decomposed into a difference of two convex functions using DC. Then a sequence of  approximations of the trailing convex function is constructed with its affine minorization (through linearizing). \emph{Second}, a quadratic problem with equality constraints is converted to an unconstrained version with slack variables, which is subsequently reconstructed by the augmented Lagrange method. \emph{Third}, the unconstrained optimization problem is solved via coordinate descent method. The detailed derivation procedures of DC, AL-CD are given in {\color{blue}Appendix D} in Supporting information. For simplicity, only the derived results are provided.

Denote $\beta_{ll'} = \beta_l - \beta_{l'}$ and define $\bmxi = (\beta_1, \cdots, \beta_p, \beta_{12}, \cdots, \beta_{1p}, \cdots, \beta_{(p-1)p})$. 
Then update $\widehat{\bmxi}^{(m,k)}$ by the following formulas, for $k = 1,2,\cdots$
\begin{itemize}
	\item Given $\hat{\beta}_{l}^{(m,k-1)}$, update $\hat{\beta}_{l}^{(m,k)} (l = 1,2,\cdots,p)$ by 
	\begin{equation*} 
	\hat{\beta}_{l}^{(m,k)} = \alpha^{-1} \gamma ,
	\end{equation*}
	where 
	$ \alpha = 2\lambda + 2\sum_{i=1}^n x_{il}^2
	+ \nu^{(k)} \left\vert l':(l,l') \in \mathcal{E}^{(m-1)} \right\vert $. 
	And $\gamma = \gamma^{*}$ if $ |\hat{\beta}_{l}^{(m-1)}| \geq \tau $; otherwise, $\gamma = \mathrm{ST}(\gamma^{*}, \frac{\lambda_1}{\tau}) $ .
	Here $\text{ST}(x, \delta) = \sign(x) (|x| - \delta)_{+}$ is the soft threshold function, and \begin{equation*}
	\gamma^{*} = 2 \sum_{i=1}^n x_{il}b_{i(-l)}^{(m,k)}
	- \sum_{(l,l') \in \mathcal{E}^{(m-1)}} \tau_{ll'}^{(k)}
	+ \nu^{(k)} \sum_{(l,l') \in \mathcal{E}^{(m-1)}}  \left(\hat{\beta}_{l'}^{(m,k)} + \hat{\beta}_{ll'}^{(m,k)} \right) , 
	\end{equation*}
	where $b_{i(-l)}^{(m,k)} = y_i - \bx^T_{i(-l)}\widehat{\bmbeta}_{(-l)}^{(m,k)}$; 
	$\bx_{i(-l)}$ is the vector $\bx_i$ without the $l$-th component, $\mathcal{E}^{(m-1)} = \{(l,l')\in \mathcal{E}, 
	|\hat{\beta}_l^{(m-1)} - \hat{\beta}_{l'}^{(m-1)}|<\tau\}\ $. 
	
	\item 
	Given $\hat{\beta}_{ll'}^{(m,k-1)}$, update $\hat{\beta}_{ll'}^{(m,k)} (1\leq l<l'\leq p)$ (with $\hat{\beta}_l^{(m,k)}$ already updated and fixed).
	Then 
	\begin{equation}\label{eq:betajj-update}
	\hat{\beta}_{ll'}^{(m,k)} = 
	\begin{cases}
	\frac{1}{\nu^{(k)}} 
	\mathrm{ST}\left( \tau_{ll'}^{(k)} + \nu^{(k)}(\hat{\beta}_l^{(m,k)} - \hat{\beta}_{l'}^{(m,k)}), \frac{\lambda_2}{\tau} \right)
	& \text{ if }(l,l') \in \mathcal{E}^{(m-1)} ,  \\
	\hat{\beta}_{ll'}^{(m-1)}  & \text{ if }(l,l') \not\in \mathcal{E}^{(m-1)} .
	\end{cases}
	\end{equation}
\end{itemize}
The process of coordinate descent iterates until convergence, satisfies the termination condition $\|\widehat{\bmbeta}^{(m,k)} -  \widehat{\bmbeta}^{(m,k-1)}\|_{\infty} \leq \delta^{*}$ (e.g. $\delta^{*} = 10^{-5}$). Hence, $\widehat{\bmbeta}^{(m)} = \widehat{\bmbeta}^{(m, t^{*})} $, where $t^{*}$ denotes the iteration at termination. 
Specially, we take $\rho=1.05, \nu=1, \delta^{*}=10^{-5}$ in the simulations.

\subsection{Convergence and Computational Complexity}
The convergence of the algorithm essentially follows the standard result. Note that we have a closed-form solution of $\bA$ when fixing $\bB$. Since the truncated penalties are not convex in $\bB$, and thus the objective function is not convex in $\bB$ when fixing $\bA$, and that is when the difference-of-convex function kicks in to convert the non-convex function to the difference of two convex functions. 
When solving the problem~\eqref{eq:givenA-solveB}, the proposed algorithm could potentially lead to a local optimum as the objective function of estimating $\bB$ when fixing $\bA$ in~\eqref{eq:givenA-solveB} is non-convex. 
But the objective function with linear constraints in AL-CD procedure obtained from the local linear approximation is differentiable everywhere, and thus the convergence of coordinate descent is guaranteed. Therefore, it is only necessary to ensure that each step is guaranteed to converge.
In Step 3, the optimized objective function is~\eqref{eq:givenB-solveA}, and we can obtain the exact solution in closed form. 
In Step 2, we solve the optimization problem~\eqref{eq:givenA-solveB} iteratively.
 The convergence of the integrated algorithm for the subproblem of estimating $\bB$ when fixing $\bA$ is given in~\autoref{lemma:converge}. Denote 
$$ S(\bmbeta)= \sum_{i=1}^n(y_i - \bx_i^T \bmbeta)^2
	+ \lambda \sum_{l=1}^{p}\beta_l^2
	+ \lambda_1 \sum_{l=1}^{p}\min \left(\frac{|\beta_l|}{\tau}, 1\right) 
	+ \lambda_2 \sum_{l < l':(l,l') \in \mathcal{E}} \min \left(\frac{|\beta_{l} - \beta_{l'}|}{\tau}, 1\right).$$

\begin{lemma}\label{lemma:converge}
	The proposed algorithm for estimation of $\bB$ given $\bA$ converges. That is 
	\begin{equation*}
	S(\widehat{\bmbeta}^{(m)}) \rightarrow c, \text{ as } m \rightarrow \infty, 
	\end{equation*}
	where $c$ is a constant value, and $m$ is the number of iterations of the integrated algorithm for problem~\eqref{eq:LS-FGFS-general}.
\end{lemma}
The~\autoref{lemma:converge} above guarantees the convergence of the algorithm for estimation of $\bB$ given $\bA$ theoretically, which is analogous to Theorem 1 in~\citet{shen2012simultaneous} and Theorem 3 in~\citet{qin2020high}. Thus we omit the proof.
It is crucial to pick a suitable initial value $\widehat{\bmbeta}^{(0)}$. 
Since~\eqref{eq:givenA-solveB} is a regression problem, possible candidate initial values are those estimated by any regression solver, such as \texttt{glmnet} in R and \texttt{sklearn} in python.

As for the computational complexity, the coordinate descent updating involves calculating of $\sum_{i=1}^{n}x_{il}^2$ and $\sum_{i=1}^{n} x_{il} b_{i(-l)}$, which requires $O(np^2k)$ operations. The construction of $\mathcal{E}$  requires $O(kp^2)$ operations. Therefore, each update in updating $\bB$ is of order $O(np^2k)$. The estimation of $\bA$ by solving an SVD needs $O(pk^2)$ operations. The total computational cost is $O(np^2k) + O(pk^2)$. 

\subsection{The Selection of Tuning Parameters}\label{sec:bic}
The cross-validation (CV) can always be one way to select the optimal values, but it is computationally expensive.
Here the Bayesian information criterion
(BIC) is employed as the approach for tuning parameter selection, which we use in simulations in ~\autoref{sec:simulation}.
In general, solutions from cross-validation and BIC
are comparable. We select the model that has the minimum BIC value when using such criteria.
Our proposed method has four tuning parameters, $\lambda, \lambda_1, \lambda_2, \tau$. 
Let $\phi$ denote the parameters that need to be tuned or selected in the candidate model. Then $\phi = (\lambda,\lambda_1,\lambda_2, \tau)$ for our proposed FGSPCA, $\phi = (\lambda,\lambda_1,\lambda_2)$ for SFPCA~\citep{guo2010principal}, and $\phi = (\lambda,\lambda_1)$ for the SPCA~\citep{zou2006sparse}.
Let $\bA^{\phi} = [\bmalpha_{1}^{\phi}, \cdots,\bmalpha_{k}^{\phi}]$, and $\bB^{\phi} = [\bmbeta_{1}^{\phi}, \cdots,\bmbeta_{k}^{\phi}]$,
be the estimates of $\bA$ and $\bB$ in~\eqref{eq:FGSPCA-crit} based on tuning parameters $\phi$. 

We define the BIC criterion of PCA variants as follows,
\begin{equation}\label{eq:bic-var}
{\rm BIC}(\phi) 
= n \log\left\{\normsFrob{\bX - \bX\bB^{\phi}(\bA^{\phi})^T}^2 / n \right\}
+ \log(n) \cdot df \ ,
\end{equation}
where $df$ represents the degree of freedom, denoted $df^{\rm SPCA}$ for SPCA~\citep{zou2006sparse}, $df^{\rm SFPCA}$ $df^{\rm SPCA}$ for SFPCA~\citep{guo2010principal}, and $df^{\rm FGSPCA}$ for FGSPCA. 
Specially, $df^{\rm SPCA}$ is defined as the number of all nonzero elements in $\bB^{\phi}$, $df^{\rm SFPCA}$ and $df^{\rm FGSPCA}$ are defined as the number of all non-distinct groups in $\bB^{\phi}$. The definitions are similar to $df$ defined for Lasso and fused Lasso~\citep{zou2007degrees, tibshirani2005sparsity}. 
Intuitively, the involvement of the truncated parameter $\tau$ makes more complex the method and the parameter tuning process. However, empirical studies show that the involvement of the truncated parameter $\tau$ establishes a trade-off between $\tau$ and $(\lambda_1, \lambda_2)$, reducing the sensitivity of the tuning of $\lambda_1$ and $\lambda_2$.

\section{Experiments}\label{sec:simulation}

\paragraph{Adjusted Variance}
Denote $\widehat{\bZ}=[\hat{Z}_1, \cdots, \hat{Z}_k]$ the modified PCs. Due to the grouping and sparsity constraints, $\hat{Z}_{k}$ is no longer orthogonal to $\hat{Z}_{i}, i=1,\cdots, k-1$. Instead, they are correlated with each other. Thus, we remove from $\hat{Z}_{k}$ the correlation effect of $\hat{Z}_{i}, i=1,\cdots, k-1$ using regression projection. 
The definition of the adjusted variance is adopted from~\citet{zou2006sparse}, which is computed based on the QR decomposition. Suppose $\widehat{\bZ}=\mathbf{Q}\mathbf{R}$, where $\mathbf{Q}$ is orthonormal and $\mathbf{R}$ is upper triangular. The adjusted variance of the $j$-th PC is $R_{jj}^2$. The explained total variance is the cumulative adjusted variance, which is defined as $\sum_{j=1}^{k} R_{jj}^2$.
\subsection{Pitprops Data}
The pitprops data is a classic dataset widely used for PCA analysis, as it is usually difficult to show the interpretability of principal components.
In the pitprops data, there are 180 observations and 13 measured variables. It is used in ScoTLASS~\citep{jolliffe2003modified} and  SPCA~\citep{zou2006sparse}. 
As a demonstration of the performance of the FGSPCA method, especially the grouping effect and sparsity effect, we consider the first six PCs of pitprops data. 
\begin{table*}[!htb]
	\centering
	\caption{Pitprops Data: Loadings of the first six PCs by SPCA and FGSPCA. The ``No. Groups'' shows the number of groups of loadings where a small number indicates strong grouping effect with similar loadings collapsing into groups. The `No. Nonzeros' is the number of non-zero loadings, which indicates the sparsity, the smaller the more sparse. 
		 `Adj.V (\%)' is the proportions of adjusted variance, and `CV (\%)' is the proportions of cumulative adjusted variance.} 
	\label{tab:FGSPCA}
	\resizebox{\textwidth}{!}{
		\begin{tabular}{lrrrrrr lrrrrrr}
			\hline \hline \\ [-1ex]
			{} & \multicolumn{6}{c}{SPCA} & & \multicolumn{6}{c}{FGSPCA} \\[0.1ex]
			\cline{2-7} \cline{9-14} \\[0.1ex]
			Variable & PC1 & PC2 & PC3 & PC4 & PC5 & PC6 & & PC1 & PC2 & PC3 & PC4 & PC5 & PC6 \\[1ex]
			\hline \hline \\[-1ex]
			topdiam & $-0.477$ & ~ & ~ & ~ & ~ & ~ & ~~ & $-0.408$ & ~ & ~ & ~ & ~ & ~ \\ 
			length & $-0.476$ & ~ & ~ & ~ & ~ & ~ & ~~ & $-0.408$ & ~ & ~ & ~ & ~ & ~ \\ 
			moist & ~ & 0.785 & ~ & ~ & ~ & ~ & ~~ & ~ & 0.707 & ~ & ~ & ~ & ~ \\ 
			testsg & ~ & 0.619 & ~ & ~ & ~ & ~ & ~~ & ~ & 0.707 & ~ & ~ & ~ & ~ \\ 
			ovensg & 0.177 & ~ & 0.641 & ~ & ~ & ~ & ~~ & ~ & ~ & 0.577 & ~ & ~ & ~ \\ 
			ringtop & ~ & ~ & 0.589 & ~ & ~ & ~ & ~~ & ~ & ~ & 0.577 & ~ & ~ & ~ \\ 
			ringbut & $-0.250$ & ~ & 0.492 & ~ & ~ & ~ & ~~ & $-0.408$ & ~ & 0.577 & ~ & ~ & ~ \\ 
			bowmax & $-0.344$ & $-0.021$ & ~ & ~ & ~ & ~ & ~~ & $-0.408$ & ~ & ~ & ~ & ~ & ~ \\ 
			bowdist & $-0.416$ & ~ & ~ & ~ & ~ & ~ & ~~ & $-0.408$ & ~ & ~ & ~ & ~ & ~ \\ 
			whorls & $-0.400$ & ~ & ~ & ~ & ~ & ~ & ~~ & $-0.408$ & ~ & ~ & ~ & ~ & ~ \\ 
			clear & ~ & ~ & ~ & $-1$ & ~ & ~ & ~~ & ~ & ~ & ~ & $-1$ & ~ & ~ \\ 
			knots & ~ & 0.013 & ~ & ~ & $-1$ & ~ & ~~ & ~ & ~ & ~ & ~ & $-1$ & ~ \\ 
			diaknot & ~ & ~ & $-0.016$ & ~ & ~ &  1 & ~~ & ~ & ~ & ~ & ~ & ~ &  1 \\ 
			No. Groups & 7 & 4 & 4 &  1 &  1 &  1 & ~~ &  1 &  1 &  1 &  1 &  1 &  1 \\ 
			No. Nonzeroes & 7 & 4 & 4 &  1 &  1 &  1 & ~~ & 6 & 2 & 3 &  1 &  1 &  1 \\ 
			Variance (\%) & 28.011 & 14.368 & 15 & 7.692 & 7.692 & 7.692 & ~~ & 28.797 & 14.477 & 15.246 & 7.692 & 7.692 & 7.692 \\ 
			Adj.V (\%) & 28.035 & 13.966 & 13.298 & 7.445 & 6.802 & 6.227 & ~~ & 28.797 & 14.099 & 11.617 & 7.442 & 6.769 & 6.233 \\ 
			CV (\%) & 28.035 & 42 & 55.299 & 62.744 & 69.546 & 75.773 & ~~ & 28.797 & 42.896 & 54.513 & 61.955 & 68.724 & 74.957 \\
			\hline \hline \\[-1ex]
		\end{tabular}
	}
\end{table*}

\begin{figure*}[htb]
	\centering
	\subfloat[][\centering PEV (Percentage of Explained Variance)]{
		\includegraphics[width=0.5\textwidth]{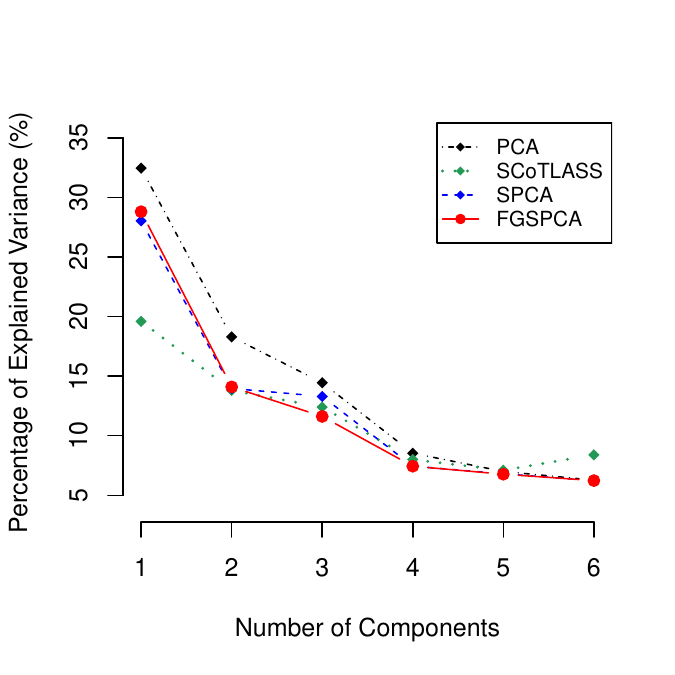} 
	}
	\subfloat[][\centering Cumulative Variance]{
		\includegraphics[width=0.5\textwidth]{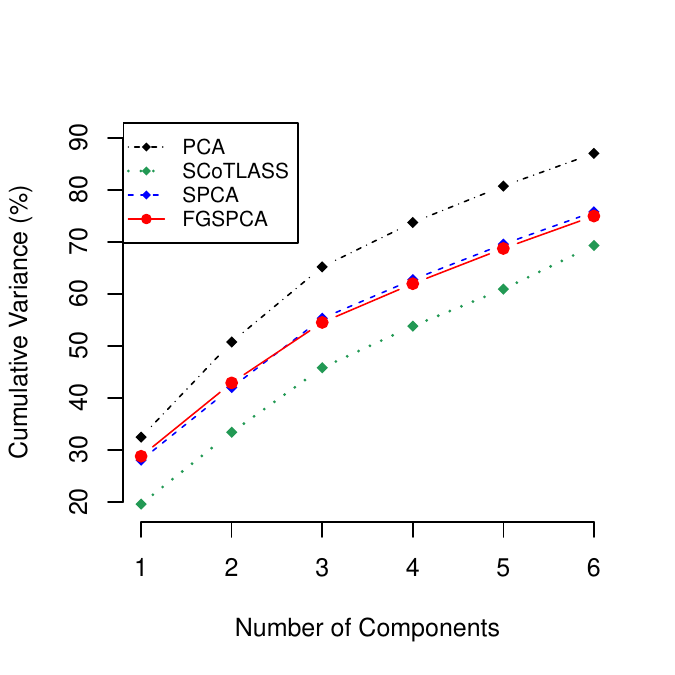} 
	}%
	\caption{Comparison of PEV, Cumulative Variance of Different Dimension Reduction Methods (PCA, SCoTLASS, SPCA, and \texttt{FGSPCA}) on the pitprops dataset.
		PCA: ordinary PCA based on SVD~\citep{jolliffe1986principal},
		SCoTLASS: modified PCA~\citep{jolliffe2003modified},
		SPCA: sparse PCA~\citep{zou2006sparse}, and FGSPCA: our new method of feature grouping and sparse PCA. }
	\label{fig:pitpev}
\end{figure*}
\autoref{tab:FGSPCA} shows the sparse loadings and the corresponding variance obtained by SPCA~\citep{zou2006sparse} and FGSPCA.
As can be seen from~\autoref{tab:FGSPCA}, both SPCA and FGSPCA show strong sparsity effects with respect to the number of zero loadings. On the other hand, FGSPCA has a strong grouping effect in terms of the number of loading groups, while SPCA has a weaker grouping effect compared to FGSPCA. Interestingly, through the grouping effect introduced in FGSPCA, FGSPCA shows a stronger sparsity compared to SPCA with respect to the number of zeroes. 
In detail, for the first PC obtained by FGSPCA, the loadings belong to two distinct groups with nonzero values and one sparse-group with zero values. Furthermore, these groups are learned automatically from the FGSPCA model rather than from prior knowledge. The grouping effect among loadings further improves the interpretability of the PCA.

It can be seen from~\autoref{fig:pitpev} that,
the first six PCs obtained by FGSPCA and SPCA account for almost the same amount of total variance, $74.96\%$ for FGSPCA and $75.77\%$ for SPCA respectively, which is much larger compared with SCoTLASS ($69.3\%$). The significant improvement in the total variance explained by FGSPCA and SPCA may result from the sparse structure on the loadings, since the derived PCs obtained by SCoTLASS are not sparse enough as analyzes in~\cite{zou2006sparse}.
\subsection{Synthetic Data}
\paragraph{Simulation 1 }
We adopt the same synthetic example settings as~\citet{zou2006sparse}. The generating mechanism of the synthetic data consists of three \emph{hidden factors}, i.e., 
\begin{equation}\label{eq:hidden3}
\begin{split}
V_1 & \sim N(0,290), \\
V_2 & \sim N(0,300), \\
V_3 & = -0.3V_1 + 0.925V_2 + \epsilon, \quad \epsilon \sim N(0,1) \ ,
\end{split}
\end{equation}
where $V_1, V_2, \epsilon$ are independent.
Next, $10$ observable variables are constructed as follows,
\begin{align*}
X_j = 
\begin{cases}
V_1 + \varepsilon_{j}, & 1 \leq j \leq 4, \\
V_2+\varepsilon_{j}, & 5 \leq j \leq 8, \\
V_3+\varepsilon_{j}, & j = 9, 10, 
\end{cases}
\end{align*}
where $\epsilon_j,~j=1,\ldots,10$, are independent and identically distributed (i.i.d) with $N(0,1)$.
Note that the variances of the three \emph{hidden factors} are $290, 300$, and $ 282.7875$ respectively.
Note that by the data generating mechanism, the variables $X_1$ to $X_4$ form a block/group with a constant weight (`block 1'), while variables $X_5$ to $X_8$ and $X_9, X_{10}$ form another two blocks, `block 2' and `block 3', respectively. Since `block 2' and `block 3' are highly correlated, thus they can be merged into one group, say `BLOCK 0'. 
Ideally, a sparse first derived PC1 should recover `BLOCK 0' of the hidden factor $V_2$ using $X_5, X_6, X_7, X_8, X_9, X_{10}$ with equal loadings, while a sparse second derived PC2 should pick up $X_1, X_2, X_3, X_4$ to recover `block 1' of the hidden factor $V_1$ with the same weights, since the variance of $V_2$ is larger than that of $V_1$. 

\citet{zou2006sparse} computed sparse PCA using the true covariance matrix as the data generating mechanism is known and the true covariance matrix of the ten observable variables $\{X_1, \cdots, X_{10}\}$ can be easily calculated.  In our simulation, we adopt the same setting procedure as in~\cite{guo2010principal} to generate data $X_{n\times p}$ with $n=50$ according to the above data generating mechanism and repeated the simulation 50 times. 
And we perform the ordinary PCA, SPCA (Sparse PCA), ST (Simple Thresholding) and FGSPCA on $X$. 
PC loadings from ordinary PCA, SPCA, ST and FGSPCA are reported in~\autoref{tab:syn-hidden-factors-pev}.
	
	\begin{table}[ht]
		\centering
		\caption{Synthetic example with three \emph{hidden factors}: Loadings of the first three/two principal components by PCA, SPCA (Sparse PCA), ST (Simple Thresholding) and FGSPCA, as well as the number of groups, the number of nonzeroes and variance. `Adj.V (\%)' is the proportions of adjusted variance, and `CV (\%)' is the proportions of cumulative adjusted variance.} 
		\label{tab:syn-hidden-factors-pev}
		\resizebox{\textwidth}{!}{
			\begin{tabular}{lrrrlrrlrrlrr}
				\toprule
				{} & \multicolumn{3}{c}{PCA} & & \multicolumn{2}{c}{SPCA} & & 
				\multicolumn{2}{c}{ST} & & \multicolumn{2}{c}{FGSPCA} \\[0.1ex]
				\cline{2-4} \cline{6-7} \cline{9-10} \cline{12-13} \\[-1ex]
				Variable & PC1 & PC2 & PC3 &~& PC1 & PC2 &~& PC1 & PC2 &~& PC1 & PC2 \\
				\midrule
				X1 & 0.116 & 0.479 & 0.062 & ~ & ~ & 0.5 & ~ & ~ & $-0.5$ & ~ & ~ & 0.5 \\ 
				X2 & 0.116 & 0.479 & 0.059 & ~ & ~ & 0.5 & ~ & ~ & $-0.5$ & ~ & ~ & 0.5 \\ 
				X3 & 0.116 & 0.479 & 0.114 & ~ & ~ & 0.5 & ~ & ~ & $-0.5$ & ~ & ~ & 0.5 \\ 
				X4 & 0.116 & 0.479 & 0.114 & ~ & ~ & 0.5 & ~ & ~ & $-0.5$ & ~ & ~ & 0.5 \\ 
				X5 & $-0.395$ & 0.145 & $-0.269$ & ~ & $-0.5$ & ~ & ~ & ~ & ~ & ~ & $-0.408$ & ~ \\ 
				X6 & $-0.395$ & 0.145 & $-0.269$ & ~ & $-0.5$ & ~ & ~ & ~ & ~ & ~ & $-0.408$ & ~ \\ 
				X7 & $-0.395$ & 0.145 & $-0.269$ & ~ & $-0.5$ & ~ & ~ & $-0.497$ & ~ & ~ & $-0.408$ & ~ \\ 
				X8 & $-0.395$ & 0.145 & $-0.269$ & ~ & $-0.5$ & ~ & ~ & $-0.497$ & ~ & ~ & $-0.408$ & ~ \\ 
				X9 & $-0.401$ & $-0.010$ & 0.582 & ~ & ~ & ~ & ~ & $-0.503$ & ~ & ~ & $-0.408$ & ~ \\ 
				X10 & $-0.401$ & $-0.010$ & 0.582 & ~ & ~ & ~ & ~ & $-0.503$ & ~ & ~ & $-0.408$ & ~ \\ 
				No. Groups & 3  & 3  & 5  & ~ & 1  & 1  & ~ & 2  & 1  & ~ & 1  & 1  \\ 
				No. Nonzeroes & 10  & 10  & 10  & ~ & 4  & 4  & ~ & 4  & 4  & ~ & 6  & 4  \\ 
				Variance (\%) & 69.64 & 30.36 & - & ~ & 41.02 & 39.65 & ~ & 38.88 & 39.65 & ~ & 59.01 & 39.65 \\ 
				Adj.V (\%) & - & - & - & ~ & 41.02 & 39.65 & ~ & 38.88 & 38.73 & ~ & \textbf{59.08} & \textbf{39.25} \\ 
				CV (\%) & 69.64 & 100  & 100  & ~ & 41.02 & 80.67 & ~ & 38.88 & 77.61 & ~ & \textbf{59.08} & \textbf{98.33} \\ 
				\bottomrule
			\end{tabular}
		}
	\end{table}
\autoref{tab:syn-hidden-factors-pev} lists three PCs of the ordinary PCA. It shows that the first two PCs account for $100\%$ of the total explained variance, suggesting that other dimension reduction methods can consider only the first two derived PCs.
	Results on~\autoref{tab:syn-hidden-factors-pev} show that all the methods (SPCA, ST, FGSPCA) can perfectly recover `block 1' with hidden factor $V_1$ using the derived PC2. 
	However, as for the first derived PC1, SPCA recovers the hidden factor $V_2$ only using $X_5, X_6, X_7, X_8$ without $X_9, X_{10}$, as the weights on $X_9, X_{10}$ are zeroes.
	The ST method recovers the hidden factor $V_2$ using $X_7, X_8, X_9, X_{10}$ which is far from being correct by imposing zero weights on $X_5, X_6$. 
	FGSPCA perfectly recovers the hidden factor $V_2$ using $X_5, X_6, X_7, X_8, X_9, X_{10}$ with the same weights, which is consistent with the ideal results analyzed above.

\begin{figure*}[htb]
	\centering
	\renewcommand{\arraystretch}{0}
	\setlength{\tabcolsep}{0pt}
	\begin{tabular}{cc}
		\includegraphics[width=0.5\linewidth]{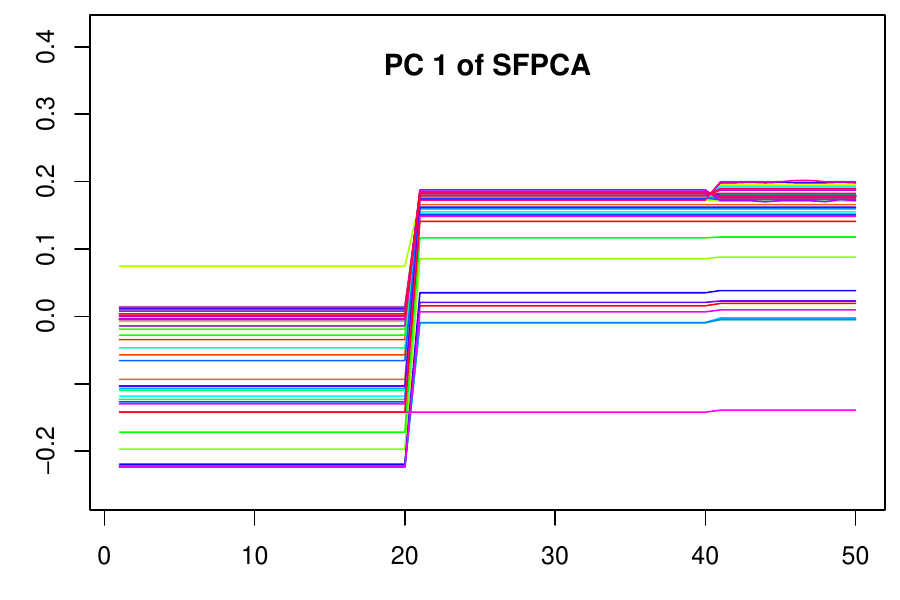} &
		\includegraphics[width=0.5\linewidth]{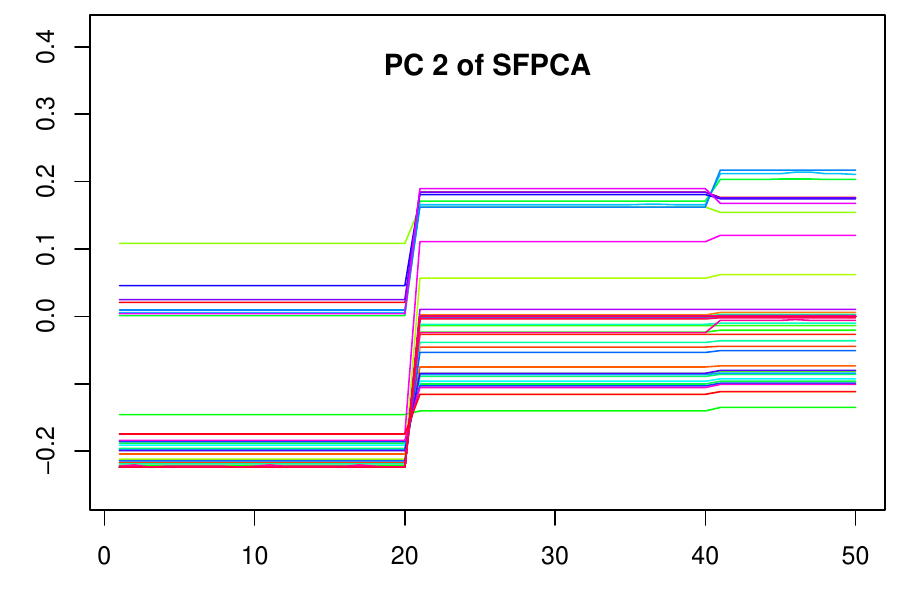} \\
		\includegraphics[width=0.5\linewidth]{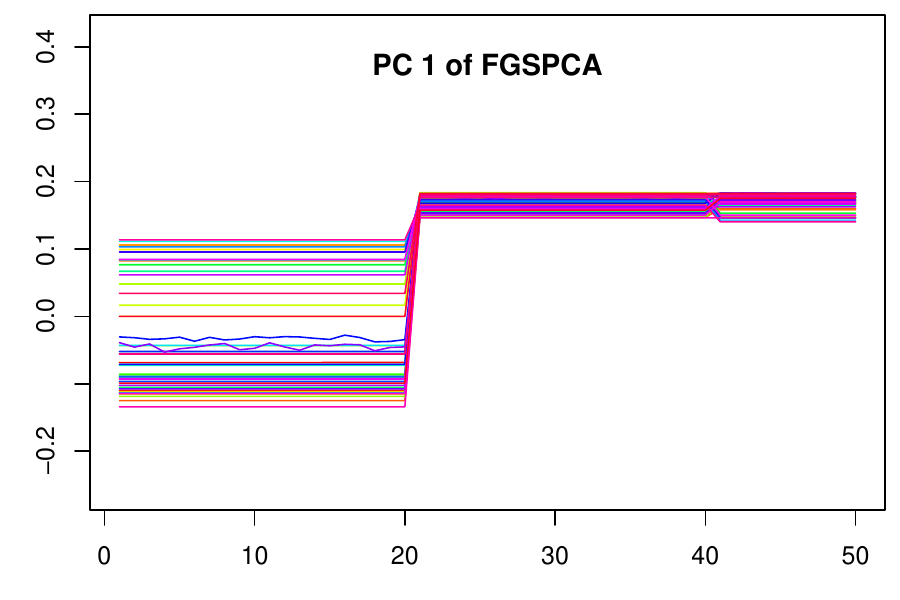} &
		\includegraphics[width=0.5\linewidth]{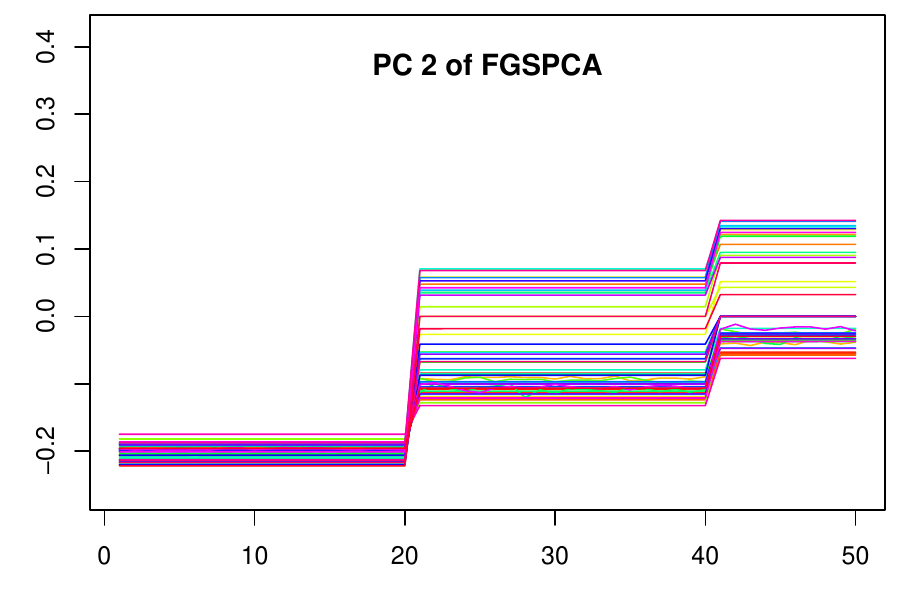}
	\end{tabular}%
	\caption{Factor loadings of the first (left column) and second (right column) PC vectors estimated by SFPCA~\citep{guo2010principal} (first row), and our proposed FGSPCA (second row). 
		The horizontal axis is the variables and the vertical axis is the value of the
		loadings. Each colored curve represents the PC vector in one replication. 
	}
	\label{fig:PC_loadings}
\end{figure*}

	The results of variance from~\autoref{tab:syn-hidden-factors-pev} show that the total variance explained by the first two PCs is $98.33\%$ for FGSPCA and $80.67\%$ for SPCA, a great improvement of $17.66\%$ due to the grouping effect of FGSPCA. 
	Moreover, compared with ordinary PCA ($100\%$ explained total variance), FGSPCA is only $1.67\%$ less with respect to the total variance explained. Most importantly, FGSPCA achieves a remarkable improvement in the interpretability of PCs with the same value, which is the grouping effect. 
	\paragraph{Simulation 2 } 
	In this example, we consider a high dimensional version ($p > n$) of Simulation 1. We define
	\begin{align*}
	X_j = 
	\begin{cases}
	V_1 + \varepsilon_{j}, & 1 \leq j \leq 20, \\
	V_2+\varepsilon_{j}, & 21 \leq j \leq 40, \\
	V_3+\varepsilon_{j}, & 41 \leq j \leq 50 \ , 
	\end{cases}
	\end{align*}
	where $\epsilon_j,~j=1,\ldots,50$, are i.i.d. $N(0, 1)$. We generate a data matrix $X_{n\times p}$ with $n=20$, and we conduct 50 repetitions. 
	The estimated loadings are illustrated in~\autoref{fig:PC_loadings}.
	Results show that SFPCA and FGSPCA produce similar sparse structures in the loadings. However, compared with the `scattered' loadings from SFPCA, the loadings estimated by FGSPCA are smooth and easier for interpretation.


\section{Discussion and Extension}\label{sec:discussion}

One limitation of FGSPCA is that it uses non-convex regularizers, neither smooth nor differentiable. Recent research work~\citep{birgin2021constrained,zhang2020adaptive,wen2018survey} has shown better denoising advantages of non-convex regularizers over convex ones. 
However, when solving the subproblem~\eqref{eq:givenA-solveB} with non-convex penalties, the proposed method could potentially lead to a local optimum, as the objective function in~\eqref{eq:givenA-solveB} is non-convex. 
As is pointed out in~\citet{wen2018survey},
the performance of non-convex optimization problems is usually closely related to the initialization, which are inherent drawbacks of non-convex optimization problems.
Hence, it is desirable to pick a suitable initial value of $\widehat{\bmbeta}^{(0)}$. 
Since each subprpblem~\eqref{eq:givenA-solveB} is a classical regression problem, possible candidate initial values are those estimated by any regression solver, such as the R package \texttt{glmnet}~\citep{friedman2009glmnet} and the python \texttt{sklearn}.
For simplicity, we use the result of SVD as the initialization in this paper.

The FGSPCA can be easily extended to the case with non-negative loadings, namely nnFGSPCA. In light of the work in~\citet{qin2020high}, we incorporate another regularization term, $p_3(\bmbeta) = \sum_{l=1}^p (\min\{\beta_l, 0\})^2$ that characterizes the non-negativity, into the objective function. The optimization problem of nnFGSPCA becomes,
\begin{equation}\label{eq:nnFGSPCA}
\begin{split}
\min_{\bA, \bB} &
\sum_{i=1}^{n} \|\bx_i - \bA\bB^T \bx_i\|_2^2 
+ \lambda \sum_{j=1}^{k} \|\bmbeta_j\|_2^2  
+ \lambda_{1} \sum_{j=1}^{k} p_1(\bmbeta_j) 
+ \lambda_2 \sum_{j=1}^{k} p_2(\bmbeta_j) 
+ \lambda_3 \sum_{j=1}^{k} p_3(\bmbeta_j) \ , \\
& \text{ s.t. } \bA^T \bA = \bI_{k\times k}, 
\end{split}
\end{equation}
The nnFGSPCA can be easily solved using similar techniques (See {\color{blue}{Appendix E}} in Supporting information for details).
\section{Conclusion}\label{sec:conclusion}
In this paper, we propose the FGSPCA method to produce modified principal components
by considering additional grouping structures where the loadings share similar coefficients (i.e., feature grouping), besides a special group with all coefficients being zero (i.e., feature selection).
The proposed FGSPCA method can perform simultaneous feature clustering/grouping and feature selection by imposing the non-convex regularization with naturally adjustable sparsity and grouping effect. 
Therefore, the model learns the grouping structure rather than from given prior information.
Efficient algorithms are designed and experiment results show that the proposed FGSPCA benefits from the grouping effect compared with methods without grouping effect.





\subsection*{Conflict of interest}

The authors declare no potential conflict of interests.

\section*{Supporting information}

Additional information for this article is available. It contains the proofs of~\autoref{lemma:one-pca}, Structured Sparsity, the Procrustes Problem and the extension to nnFSGPCA. 
An R implementation of FGSPCA can be found on github~\href{ https://github.com/higeeks/FGSPCA}{https://github.com/higeeks/FGSPCA}.




\nocite{*}
\bibliography{references}%

\articletype{Supporting information}%


\title{Supporting information of ` Feature Grouping and Sparse Principal Component Analysis with Truncated Regularization'
}

\author[1]{Haiyan Jiang}

\author[2]{Shanshan Qin*}

\author[3]{Oscar Hernan Madrid Padilla}

\authormark{Jiang  \textsc{et al}}

\address[1]{\orgdiv{School of Statistics and Data Science}, \orgname{Nankai University}, \orgaddress{\state{Tianjin}, \country{China}}. \texttt{Email: jianghaiyan.cn@gmail.com} }

\address[2]{\orgdiv{School of Statistics}, \orgname{ Tianjin University of Finance and Economics}, \orgaddress{\state{Tianjin}, \country{China}}. \texttt{Email: qinsslzu@gmail.com}}

\address[3]{\orgdiv{Department of Statistics}, \orgname{University of California}, \orgaddress{\state{California}, \country{USA}}. \\ \texttt{Email: oscar.madrid@stat.ucla.edu}}

\corres{*Shanshan Qin. \email{qinsslzu@gmail.com}}





\appendix
\section{Proof of LEMMA 1}\label{append:A}
\begin{lemma}\label{lemma:appendA}
	Consider the ridge regression criterion,
	\[ 
	J_{\lambda}(\bmbeta)= \|\by - \bX\bmbeta\|_2^2 + \lambda \|\bmbeta\|_2^2 \ ,
	\]
	Denote the solution of ridge regression 
	${ \widehat{\bmbeta} = \argmin_{\bmbeta} J_{\lambda}(\bmbeta) 
		= (\bX^T\bX + \lambda\bI)^{-1} \bX^T\by .} $
	Then 
	\begin{equation*}
	J_{\lambda}(\widehat{\bmbeta})
	= \by^T (\bI - \bH_{\lambda}) \by ,
	\end{equation*}
	where 
	\[
	\bH_{\lambda} = 
	\bX (\bX^T\bX + \lambda\bI)^{-1} \bX^T .
	\]
\end{lemma}

\paragraph{Proof of Lemma 1.}
We use the notation $\bA_{p\times k}=[\bmalpha_1, \cdots, \bmalpha_k]$ and $\bB_{p\times k}=[\bmbeta_1, \cdots, \bmbeta_k]$. 
Let
\[
J_{\lambda}(\bA, \bB)
=  \sum_{i=1}^{n} \|\bx_i - \bA\bB^T\bx_i\|_2^2 + \lambda\|\bmbeta\|_2^2 \ .
\]
Define $\bI_{p\times p} = [\bA \ \bA_{\perp}]$. And $\bA \in \mathbb{R}^{p \times k}$, $\bA_{\perp} \in \mathbb{R}^{p \times (p-k)}$, $\bB \in \mathbb{R}^{p\times k}$. With the orthogonal constraint, $\bA^{T}\bA = \bI_{k \times k}$, $\bB^{T}\bB = \bI_{k \times k}$, and constraint  $\bA^{T}\bA_{\perp} = \mathbf{0}_{k\times (p-k)}$, we have
\begin{equation}\label{eq:A1}
\begin{split}
\sum_{i=1}^{n} \|\bx_i - \bA\bB^T\bx_i\|_2^2 
& = \normsFrob{ \bX - \bX \bB \bA^{T} } ^{2} \\
& = \normsFrob{ \bX[\bA \ \bA_{\perp}] - \bX \bB \bA^{T}[\bA \ \bA_{\perp}] }^{2} \\
& = \normsFrob{ [\bX \bA \quad \bX\bA_{\perp}] - [\bX \bB \bA^{T}\bA \quad \bX \bB \bA^{T}\bA_{\perp}] }^{2} \\
& = \normsFrob{ [(\bX\bA - \bX\bB) \quad \bX\bA_{\perp}] }^2 \quad (\bA^{T}\bA=\bI_{k\times k}, \bA^{T}\bA_{\perp}=0 )\\
& = \normsFrob{\bX\bA - \bX\bB }^2 + \normsFrob{\bX\bA_{\perp} }^2 \\
& = \sum_{j=1}^{k} \|\bX\bmalpha_j - \bX\bmbeta_j\|_2^2 
+ \normsFrob{\bX\bA_{\perp} }^2 \ . 
\end{split}
\end{equation}
Therefore, when $\bA$ is fixed, solving $\argmin_{\bB} J_{\lambda}(\bA, \bB)$ is equivalent to solving the series of ridge regressions
\[
\argmin_{\{\bmbeta_j\}_{j=1}^{(k)}}
\sum_{j=1}^{k} \left\{ \|\bX\bmalpha_j - \bX\bmbeta_j\|_2^2 + \lambda \|\bmbeta_j\|_2^2 \right\} .
\]
Denote the solutions 
\begin{equation}\label{eq:A2}
\widehat{\bB} = (\bX^T\bX+\lambda\bI)^{-1}\bX^T\bX\bA .
\end{equation}
Using~\autoref{lemma:appendA},~\eqref{eq:A1} and~\eqref{eq:A2}, we have the partially optimized criterion
\begin{equation*} 
\begin{split}
J_{\lambda}(\bA, \widehat{\bB})
& = \normsFrob{ \bX\bA_{\perp} }^2 + \sum_{j=1}^{k} \left\{ \|\bX\bmalpha_j - \bX\widehat{\bmbeta}_j\|_2^2 + \lambda \|\widehat{\bmbeta}_j\|_2^2 \right\} \\
& = \normsFrob{\bX\bA_{\perp} }^2 
+ \tr\{(\bX\bA)^T(\bI-\bH_{\lambda}) (\bX\bA) \} . 
\end{split}
\end{equation*}
Note that 
\begin{equation*}
\begin{split}
\normsFrob{\bX\bA_{\perp} }^2 + \tr\{ (\bX\bA)^T(\bX\bA) \}
& = \tr\{ (\bX\bA_{\perp})^T(\bX\bA_{\perp}) \} + \tr\{ (\bX\bA)^T(\bX\bA) \} \\
& = \normsFrob{\bX\bA_{\perp} }^2 + \normsFrob{\bX\bA }^2 = \normsFrob{\bX[\bA \ \bA_{\perp}] }^2 \\
& = \normsFrob{\bX}^2 = \tr(\bX^T \bX) .
\end{split}
\end{equation*}
Then we have 
\begin{equation}\label{eq:A4}
J_{\lambda}(\bA, \widehat{\bB})
= \tr(\bX^T \bX)
+ \tr(\bA^T \bX^T\bH_{\lambda}\bX \bA) \ ,
\end{equation}
which should be minimized with respect to $\bA$ with constraint that $\bA^T\bA=\bI $. 
The solution of~\eqref{eq:A4} should be taken to the top $k$ eigenvectors of $\bX^T\bH_{\lambda}\bX $. 
If the SVD of $\bX = \bU\bD\bV^T$, we can easily get 
\[
\bX^T\bH_{\lambda}\bX = \bV \bD^2(\bD^2+\lambda\bI)^{-1}\bD^2 \bV^T, 
\]
then we have 
\[ \widehat{\bA} =\bV[, 1:k] . \]
By plugging in the SVD of $\bX$ into~\eqref{eq:A2}, each $\widehat{\bmbeta}_j$ is proportional to $V_j$ with
\[
\widehat{\bmbeta}_j = V_{j} \frac{\bD_{jj}^2}{\bD_{jj}^2 + \lambda} \propto V_j \ .
\]

\section{Structured Sparsity}\label{append:B}

We consider the structured regularization functions of variables (factors) in regression models, as variable selection and model selection are two essential issues which have been extensively studied in the framework of regression, especially in the high dimensional settings.

\paragraph{Elastic net~\citep{zou2005regularization}}
The naive elastic net criterion~\citep{zou2005regularization} is defined as 
\begin{equation}\label{e1:elasticnet}
\|Y - \bX\bmbeta \|_2^2 + \lambda_2 \|\bmbeta\|_2^2 + \lambda_1 \|\bmbeta\|_1 \ .
\end{equation}
If there is a group of variables among which the pairwise correlations are very high, the lasso tends to select only one variable from the group and does not care which one is selected. 
Unlike the lasso, the elastic net encourages a grouping effect, where strongly correlated predictors tend to be in or out of the model together.

\paragraph{Fused lasso~\citep{tibshirani2005sparsity}}

The fused lasso~\citep{tibshirani2005sparsity} is defined as follows, 
\begin{equation}\label{eq:fusedlasso}
\widehat{\bmbeta} = \argmin_{\bmbeta} \| Y-\bX\bmbeta\|_2^2, 
\quad \text{ subject to } 
\sum_{j=1}^{p} |\beta_j| \leq s_1, \quad 
\sum_{j=2}^{p} |\beta_j - \beta_{j-1}| \leq s_2 . 
\end{equation}
The first constraint induces sparsity in the coefficients; the second results in sparsity in their successive differences, i.e. local constancy of the coefficient profiles $\beta_j$ as a function of $j$. 
The fused lasso gives a way to incorporate information about spatial or temporal structure in the data.
However, it requires the features to be ordered in some meaningful way before the construction of the problem.

\paragraph{Group lasso~\citep{yuan2006model}}

Consider the general regression problem with $J$ groups/factors,
\begin{equation*}
Y = \sum_{j=1}^{J}\bX_j \bmbeta_j + \epsilon .
\end{equation*}
Here $Y \in \mathbb{R}^{n \times 1}$, $\epsilon  \sim N_{n}(0, \sigma^2 I)$, $\bX_j$ is an $n\times p_j$ matrix corresponding to the $j$-th factor and $\bmbeta_j$ is the coefficient vector of size $p_j$, $j=1,\cdots, J$.
For a vector $\eta \in \mathbb{R}^{d}$, $d \geq 1$, and a symmetric $d\times d$ positive definite matrix $\bK$, we denote $ {\|\eta \|_{\bK} = (\eta \bK \eta)^{1/2} } $ . 
Given positive definite matrices $\bK_1, \cdots, \bK_J$, the group lasso~\citep{yuan2006model} estimate is defined as the solution to the following minimization problem, 
\begin{equation}\label{eq:grouplasso}
\min_{\bmbeta} = \frac{1}{2n} \|Y - \sum_{j=1}^{J}\bX_j \bmbeta_j \|_2^2 
+ \lambda \sum_{j=1}^{J} \|\bmbeta_j\|_{\bK_j} \ . 
\end{equation}
In the group lasso problem, the non-squared Euclidean $\ell_2$-norm penalty encourages factor/group-level sparsity, where the entire group of predictors can be retained or discarded in the model. Thus the group lasso can conduct feature selection along the group level and select groups of variables. However, this kind of group-level sparsity depends on the predefined group partition.

\paragraph{Structured sparsity-inducing norms~\citep{jenatton2011structured}}

Consider the empirical risk minimization problem for linear supervised learning, with regularization by structured sparsity-inducing norms~\citep{jenatton2011structured}, 
\begin{equation}\label{eq:ssnorm}
\min_{\bw} \frac{1}{n} \sum_{i=1}^{n}\ell(y_i, \bx_i^T \bw) + \lambda \Omega(\bw) \ ,
\end{equation}
where $\lambda$ is a regularization parameter, $L(\bw) = \frac{1}{n} \sum_{i=1}^{n}\ell(y_i, \bx_i^T \bw)$ the empirical risk of a weight vector $\bw \in \mathbb{R}^{p}$, $\ell(\cdot, \cdot)$ is a loss function which is usually assumed convex and continuously differentiable with respect to the second parameter. 
The $\Omega(\bw) $ is a general family of sparsity-inducing norms that allow the penalization of subsets of variables grouped together, which is defined as follows, 
\begin{equation}\label{eq:ssnorm-pen}
\Omega(\bw) = \sum_{G \in \mathcal{G}} 
\left[ \sum_{j \in G} (d_{j}^{G})^2 |w_j|^2 \right]^{\frac{1}{2}} 
= \sum_{G \in \mathcal{G}} \norms{d^{G} \circ \bw }_2 \ .
\end{equation}
Here $(d^G)_{G\in \mathcal{G}}$ is a $|G|$-tuple of $p$-dimensional vectors such that $d_{j}^{G}>0$ if $j \in G$ and $d_{j}^{G}=0$ otherwise, 
and $G$ denotes a subset of the power set of $\{1, \cdots, p\}$ such that 
${ \sum_{G \in \mathcal{G}} = \{1, \cdots, p\} }$, that is, a spanning set of subsets of $\{1, \cdots, p\}$. 
It is possible for elements of $\mathcal{G}$ to overlap. 
This general formulation has several important sub-cases such as $\ell_2$-norm penalty, $\ell_1$-norm penalty, group $\ell_1$-norm penalty, and elastic net penalty. However, the structured sparsity-inducing regularization can only encode \emph{prior knowledge} about the expected sparsity patterns.

\section{The Procrustes Problem}



\paragraph{Lemma S1}
	Reduced Rank Procrustes Rotation. 
	$\bM_{n\times p}$ and $\bN_{n\times k}$ denote two matrices.
	Consider the constrained minimization problem
	\begin{equation}\label{eq:solveA}
	\widehat{\bA} = \argmin_{A} \normsFrob{\bM - \bN \bA^T}^2 \quad 
	\mathrm{s.t. } \quad \bA^T\bA = \bI_{k\times k} \ .
	\end{equation}
	Suppose the SVD of $\bM^T\bN$ is $\bU\bD\bV^T$, then $\widehat{\bA}=\bU\bV^T$. 

\paragraph{Proof of Lemma S1}
In the orthogonal Procrustes problem, we seek an orthornormal matrix such that 
\begin{equation*}
\bA = \argmin_{\bA} \normsFrob{\bM - \bN\bA^T }^2, \text{ s.t. } \bA^T\bA=\bI_{k\times k} .
\end{equation*}
First, we expand the matrix norm in the above objective function
\[
\normsFrob{\bM - \bN\bA^T }^2
= \tr(\bM^T\bM) + \tr(\bA\bN^T\bN\bA^T)
- 2\tr(\bM^T\bN\bA^T) \ .
\]
Since $\bA^T\bA=\bI_{k\times k}$ and $\tr(\bA\mathbf{B}) = \tr(\mathbf{B}\bA)$, then the second term becomes 
\[
\tr(\bA\bN^T\bN\bA^T) = \tr(\bN^T\bN\bA^T\bA) = \tr(\bN^T\bN) \ .
\]
The problem is equivalent to finding an orthornormal matrix $\bA$ which maximize $\tr(\bM^T\bN\bA^T) $. 
We proceed by substituting the SVD of $\bM^T\bN = \bU\bD\bV^T$ and obtain
\begin{equation*}
\tr(\bM^T\bN\bA^T) = \tr(\bU\bD\bV^T\bA^T) 
= \tr\{\bU\bD(\bA\bV)^T\} 
= \tr\{(\bA\bV)^T\bU\bD\} .
\end{equation*}
As $\bV$ is $k \times k$ orthonormal, we have $(\bA\bV)^T(\bA\bV) = \bV^T\bA^T\bA\bV = \bI_{k\times k}$ .
Note that $\bD$ is diagonal with non-negative entries, $\tr\{(\bA\bV)^T\bU\bD\}$ is maximized when the diagonal of $(\bA\bV)^T\bU$ is positive and maximized.  By Cauchy-Schwartz inequality, this is achieved when $(\bA\bV)^T = \bU^T$, and in this case the diagonal elements are all ones, $(\bA\bV)^T\bU=\bU^T\bU = \bI$ . 
Hence, an optimal solution is given by 
$\widehat{\bA} = \bU\bV^T$ .

\section{The detailed procedure to Estimation of \texorpdfstring{$\bB$}{TEXT} given \texorpdfstring{$\bA$}{TEXT}}

An integrated algorithm for the estimation of $\bB$ given $\bA$ (algorithm 1) integrates the difference-of-convex algorithm (DC), the augmented Lagrange method (AL) and coordinate descent method (CD), for efficient computation. The procedure to solve the FGS problem consists of three steps.
\paragraph{The Difference-of-Convex Algorithm (DC).}
Denote 
{$$ S(\bmbeta)= \sum_{i=1}^n(y_i - \bx_i^T \bmbeta)^2
	+ \lambda \sum_{l=1}^{p}\beta_l^2
	+ \lambda_1 \sum_{l=1}^{p}\min \left(\frac{|\beta_l|}{\tau}, 1\right) 
	+ \lambda_2 \sum_{l < l':(l,l') \in \mathcal{E}} \min \left(\frac{|\beta_{l} - \beta_{l'}|}{\tau}, 1\right).$$}
Using $\min(a, b)=a-(a-b)_{+}$, we decompose the non-convex objective function $S(\bmbeta)$ into a difference of two convex functions, $ { S(\bmbeta) = S_{1}(\bmbeta) - S_{2}(\bmbeta) }$, 
where the two convex functions are given respectively by
\begin{align*}
&S_1(\bmbeta) = \sum_{i=1}^n (y_i - \bx_i^T\bmbeta)^2 
+ \lambda \sum_{l=1}^{p} \beta_{l}^2 
+ \lambda_1 \sum_{l=1}^{p} \frac{|\beta_{l}|}{\tau} 
+ \lambda_2 \sum_{l<l':(l,l')\in \mathcal{E}}^p \frac{|\beta_{l}-\beta_{l'}|}{\tau} \ , \\
&S_2(\bmbeta) = \lambda_1 \sum_{l=1}^{p} \left(\frac{|\beta_{l}|}{\tau} - 1\right)_{+} 
+ \lambda_2 \sum_{l<l':(l,l')\in \mathcal{E}}^p \left(\frac{|\beta_{l}-\beta_{l'}|}{\tau}-1\right)_{+} .
\end{align*}
We then construct a sequence of approximations of $S_2(\bmbeta)$ iteratively. At the $m$-th iteration, we replace $S_2(\bmbeta)$ with its affine minorization at the $(m-1)$-th iteration.
Specially,
\begin{align*}
S_2^{(m)}(\bmbeta) 
& = S_2(\widehat{\bmbeta}^{(m-1)}) 
+ \langle\bmbeta - \widehat{\bmbeta}^{(m-1)}, 
\left.\partial S_2(\bmbeta)\right\vert_{\bmbeta = \widehat{\bmbeta}^{(m-1)}}\rangle \\
& = S_2(\widehat{\bmbeta}^{(m-1)}) 
+ \frac{\lambda_1}{\tau} \sum_{l=1}^p 
\mathrm{I}_{\left\{|\hat{\beta}_{l}^{(m-1)}|\geq \tau \right\} } \cdot |\beta_l|
+ \frac{\lambda_2}{\tau} \sum_{l<l':(l, l')\in \mathcal{E}} \mathrm{I}_{ \left\{ |\hat{\beta}_{l}^{(m-1)} - \hat{\beta}_{l'}^{(m-1)}|\geq \tau \right\} } 
\cdot |\beta_l - \beta_{l'}|\ .
\end{align*}
Finally, a sequence of approximations of $S(\bmbeta)$ is constructed iteratively. For the $m$-th approximation, an upper convex approximating function to $S(\bmbeta)$ can be obtained by $S^{(m)}(\bmbeta) = S_1(\bmbeta) - S_2^{(m)}(\bmbeta)$, which formulates the following subproblem:
\begin{equation}\label{eq:sub-problem}
\min_{\bmbeta} S^{(m)}(\bmbeta) = \sum_{i=1}^n (y_i - \bx_i^T\bmbeta)^2 
+ \lambda \sum_{l=1}^{p} \beta_l^2
+ \frac{\lambda_1}{\tau} \sum_{l \in \mathcal{F}^{(m-1)}} |\beta_{l}| 
+ \frac{\lambda_2}{\tau} \sum_{l<l': (l,l')\in \mathcal{E}^{(m-1)} } |\beta_l - \beta_{l'}|\ ,
\end{equation}
where 
\begin{equation}\label{eq:EF-update}
\begin{split}
&\mathcal{F}^{(m-1)} = \{l:|\hat{\beta}_l^{(m-1)}|<\tau\}\ , \\
&\mathcal{E}^{(m-1)} = \{(l,l')\in \mathcal{E}, 
|\hat{\beta}_l^{(m-1)} - \hat{\beta}_{l'}^{(m-1)}|<\tau\}\ .
\end{split}
\end{equation}

\paragraph{Augmented Lagrange Method and Coordinate Descent Method (AL-CD).}
Denote $\beta_{ll'} = \beta_l - \beta_{l'}$ and define $\bmxi = (\beta_1, \cdots, \beta_p, \beta_{12}, \cdots, \beta_{1p}, \cdots, \beta_{(p-1)p})$. The $m$-th subproblem~\eqref{eq:sub-problem} can be reformulated as an equality-constrained convex optimization problem, 
\begin{equation}\label{eq:sub-eq-const}
\begin{split}
\min_{\bmxi} & \sum_{i=1}^n (y_i - \bx_i^T\bmbeta)^2 
+ \lambda \sum_{l=1}^{p} \beta_l^2 
+ \frac{\lambda_1}{\tau} \sum_{l \in \mathcal{F}^{(m-1)}} |\beta_l| 
+ \frac{\lambda_2}{\tau} \sum_{l<l': (l,l')\in \mathcal{E}^{(m-1)}} | \beta_{ll'}| \ , \\
& \text{ subject to } \beta_{ll'} = \beta_l - \beta_{l'}, \quad  \forall l<l':(l,l')\in \mathcal{E}^{(m-1)} \ .
\end{split}
\end{equation}
For the equality-constrained problem~\eqref{eq:sub-eq-const}, we employ the augmented Lagrange method to solve its equivalent unconstrained version iteratively with respect to $k$ for the $m$-th approximation. 
For the $m$-th approximation, the augmented Lagrange method for~\eqref{eq:sub-eq-const} is
\begin{equation} \label{eq:alm}
\begin{split}
L_{\nu}^{(m)}(\bmxi, \boldsymbol{\tau}) 
= & \sum_{i=1}^n (y_i - \bx_i^T\bmbeta)^2 
+ \lambda \sum_{l=1}^{p} \beta_l^2 
+ \frac{\lambda_1}{\tau} \sum_{l \in \mathcal{F}^{(m-1)}} |\beta_l|  
+ \frac{\lambda_2}{\tau} \sum_{l<l': (l,l')\in \mathcal{E}^{(m-1)}} | \beta_{ll'}| \\
& + \sum_{l<l': (l,l')\in \mathcal{E}^{(m-1)}} \tau_{ll'}(\beta_{l} - \beta_{l'} - \beta_{ll'}) 
+ \frac{\nu}{2} \sum_{l<l': (l,l')\in \mathcal{E}^{(m-1)}} (\beta_{l} - \beta_{l'} - \beta_{ll'})^2\ .
\end{split}
\end{equation}
Here $\tau_{ll'}$ and $\nu$ are the Lagrangian multipliers for the linear constraints and for the computational acceleration, which are updated as follows, 
\begin{equation}\label{eq:uv-update}
\tau_{ll'}^{(k+1)} 
= \tau_{ll'}^{(k)} + \nu^{(k)} 
(\hat{\beta}_{l}^{(m,k)} - \hat{\beta}_{l'}^{(m,k)} - \hat{\beta}_{ll'}^{(m,k)}), \quad
\nu^{(k+1)} = \rho \nu^{(k)} \ .
\end{equation}
Here $\rho$ controls the convergence speed of the algorithm, which is chosen to be larger than $1$ (e.g. $\rho=1.05$) for acceleration of the convergence.

For the $\bmxi$ minimization step in~\eqref{eq:alm}, we use the coordinate descent methods to compute the update. Denote a solution of~\eqref{eq:alm} as $\widehat{\bmxi}^{(m,k+1)}$.
For each component of $\bmxi$, we fix the other components at their current values. Set an initial value $\widehat{\bmxi}^{(m,0)} = \widehat{\bmxi}^{(m-1)}$, where $\widehat{\bmxi}^{(m-1)}$ is the solution of the subproblem~\eqref{eq:sub-problem} for the $(m-1)$-th approximation. Then update $\widehat{\bmxi}^{(m,k)}$ by the following formulas, for $k = 1,2,\cdots$
\begin{itemize}
	\item Given $\hat{\beta}_{l}^{(m,k-1)}$, update $\hat{\beta}_{l}^{(m,k)} (l = 1,2,\cdots,p)$ by 
	\begin{equation*} 
	\hat{\beta}_{l}^{(m,k)} = \alpha^{-1} \gamma ,
	\end{equation*}
	where 
	$ \alpha = 2\lambda + 2\sum_{i=1}^n x_{il}^2
	+ \nu^{(k)} \left\vert l':(l,l') \in \mathcal{E}^{(m-1)} \right\vert $. 
	And $\gamma = \gamma^{*}$ if $ |\hat{\beta}_{l}^{(m-1)}| \geq \tau $; otherwise, $\gamma = \mathrm{ST}(\gamma^{*}, \frac{\lambda_1}{\tau}) $ .
	Here $\text{ST}(x, \delta) = \sign(x) (|x| - \delta)_{+}$ is the soft threshold function, and \begin{equation*}
	\gamma^{*} = 2 \sum_{i=1}^n x_{il}b_{i(-l)}^{(m,k)}
	- \sum_{(l,l') \in \mathcal{E}^{(m-1)}} \tau_{ll'}^{(k)}
	+ \nu^{(k)} \sum_{(l,l') \in \mathcal{E}^{(m-1)}}  \left(\hat{\beta}_{l'}^{(m,k)} + \hat{\beta}_{ll'}^{(m,k)} \right) , 
	\end{equation*}
	where $b_{i(-l)}^{(m,k)} = y_i - \bx^T_{i(-l)}\widehat{\bmbeta}_{(-l)}^{(m,k)}$; 
	$\bx_{i(-l)}$ is the vector $\bx_i$ without the $l$-th component. 
	
	\item 
	Given $\hat{\beta}_{ll'}^{(m,k-1)}$, update $\hat{\beta}_{ll'}^{(m,k)} (1\leq l<l'\leq p)$ (with $\hat{\beta}_l^{(m,k)}$ already updated and fixed).
	Then 
	\begin{equation*} 
	\hat{\beta}_{ll'}^{(m,k)} = 
	\begin{cases}
	\frac{1}{\nu^{(k)}} 
	\mathrm{ST}\left( \tau_{ll'}^{(k)} + \nu^{(k)}(\hat{\beta}_l^{(m,k)} - \hat{\beta}_{l'}^{(m,k)}), \frac{\lambda_2}{\tau} \right)
	& \text{ if }(l,l') \in \mathcal{E}^{(m-1)} ,  \\
	\hat{\beta}_{ll'}^{(m-1)}  & \text{ if }(l,l') \not\in \mathcal{E}^{(m-1)} .
	\end{cases}
	\end{equation*}
\end{itemize}
The process of coordinate descent iterates until convergence, satisfies the termination condition $\|\widehat{\bmbeta}^{(m,k)} -  \widehat{\bmbeta}^{(m,k-1)}\|_{\infty} \leq \delta^{*}$ (e.g. $\delta^{*} = 10^{-5}$). Hence, $\widehat{\bmbeta}^{(m)} = \widehat{\bmbeta}^{(m, t^{*})} $, where $t^{*}$ denotes the iteration at termination. 
Specially, we take $\rho=1.05, \nu=1, \delta^{*}=10^{-5}$ in the simulations.

\section{Extension to nnFGSPCA}\label{append:E}

\paragraph{The nnFGSPCA criterion}
For the FGSPCA criterion, by adding another regularization function controlling the non-negativity of the loadings, we can obtain \textbf{the nnFGSPCA criterion}~\eqref{eq:nnFGSPCA-crit} easily,
\begin{equation}\label{eq:nnFGSPCA-crit}
\begin{split}
\min_{\bA, \bB}
& \sum_{i=1}^{n} \|\bx_i - \bA\bB^T \bx_i\|_2^2  
+ \Psi(\mathbf{B}) \ , \\
& \text{ subject to } \bA^T \bA = \bI_{k\times k} \ , 
\end{split}
\end{equation}
where 
\begin{equation}
\Psi(\mathbf{B}) = \lambda \sum_{j=1}^{k} \|\bmbeta_j\|_2^2  
+ \lambda_{1} \sum_{j=1}^{k} p_1(\bmbeta_j) 
+ \lambda_2 \sum_{j=1}^{k} p_2(\bmbeta_j)
+ \lambda_3 \sum_{j=1}^{k} p_3(\bmbeta_j)\ .
\end{equation}
Here ${p}_{1}(\bmbeta)$ and ${p}_{2}(\bmbeta)$ are the same regularization functions as that in the FGSPCA criterion, and  
${p}_{3}(\bmbeta)$ is a new regularization function controlling the non-negativity of the loadings, which takes the following penalty form,
\begin{equation}\label{eq:pen-nonnegative}
p_3(\bmbeta_j) = \sum_{l=1}^{p} 
\left[\min\left(\beta_{l(j)}, 0\right) \right]^2 \ .
\end{equation}
In order to be self-contained, we also list here the regularization functions of ${p}_{1}(\bmbeta)$ and ${p}_{2}(\bmbeta)$  
\begin{equation*}
p_1(\bmbeta_j) = \sum_{l=1}^{p} \min\left\{\frac{|\beta_{l(j)}|}{\tau}, 1\right\}, \quad
p_2(\bmbeta_j) = \sum_{l<l': (l, l') \in \mathcal{E}} \min\left\{ \frac{|\beta_{l(j)} - \beta_{l'(j)}|}{\tau}, 1\right\} \ .
\end{equation*}
The algorithm to solve the nnFGSPCA problem should be similar to the algorithms in Section 4. The procedure of updating $\bA$ is the same, only the updating of $\hat{\beta}_{l}^{(m,k)}$ is slightly different.

\paragraph{To calculate $\bB$.}
If $\bA$ is given, for each $j$, denote $\bY_j = \bX \bmalpha_j$. To estimate $\widehat{\bB} =[\widehat{\bmbeta}_1, \cdots, \widehat{\bmbeta}_k]$, the nnFGSPCA criterion is equivalent to $k$ independent non-negative feature-grouping-and-sparsity constrained regression subproblems (nnFGS) defined in the following
\begin{equation}\label{eq:nn-givenA-solveB}
\min_{\bmbeta} \left\{ S(\bmbeta) = \|\bY_j - \bX\bmbeta\|_2^2
+ \lambda \|\bmbeta\|_2^2
+ \lambda_{1} p_1(\bmbeta) + \lambda_2 p_2(\bmbeta) 
+ \lambda_{3} p_3(\bmbeta) \right\}.
\end{equation}
Each $\widehat{\bmbeta}_j = \argmin_{\bmbeta} S(\bmbeta)$ is a solution of the nnFGS problem, which can be obtained through a slightly different updating process of $\hat{\beta}_{l}^{(m, k)}$. 

Note that $p_3(\bmbeta)$ should be decomposed by the difference-of-convex programming just as $p_1(\bmbeta)$ and $p_2(\bmbeta)$ do. 
In particular, $S(\bmbeta)$ can be decomposed as follows, 
\begin{equation*}
S(\bmbeta) = S_{1}(\bmbeta) - S_{2}(\bmbeta)\ ,
\end{equation*} 
where the two convex functions $S_{1}(\bmbeta)$ and $S_{2}(\bmbeta)$ are given respectively by 
\begin{align*}
&S_1(\bmbeta) = \sum_{i=1}^n ( y_i - \bx_i^T\bmbeta)^2 
+ \lambda \sum_{l=1}^{p} \beta_{l}^2 
+ \lambda_1 \sum_{l=1}^{p} \frac{|\beta_{l}|}{\tau} 
+ \lambda_2 \sum_{l<l':(l,l')\in \mathcal{E}}^p \frac{|\beta_{l}-\beta_{l'}|}{\tau} 
+ \lambda_3 \sum_{l=1}^{p} \beta_{l}^2 \ , \\
&S_2(\bmbeta) = \lambda_1 \sum_{l=1}^{p} \left(\frac{|\beta_{l}|}{\tau} - 1\right)_{+} 
+ \lambda_2 \sum_{l<l':(l,l')\in \mathcal{E}}^p \left(\frac{|\beta_{l}-\beta_{l'}|}{\tau}-1\right)_{+} 
+ \lambda_3 \sum_{l=1}^{p} [(\beta_{l})_{+}]^2.
\end{align*}

For the $m$-th iteration, we replace $S_2(\bmbeta)$ with its affine minorization at the $(m-1)$-th iteration. 
\begin{align*}
S_2^{(m)}(\bmbeta) 
\propto & S_2(\widehat{\bmbeta}^{(m-1)}) 
+ \left\langle\bmbeta - \widehat{\bmbeta}^{(m-1)}, 
\left.\partial S_2(\bmbeta)\right\vert_{\bmbeta = \widehat{\bmbeta}^{(m-1)}} \right\rangle\\
\propto & S_2(\widehat{\bmbeta}^{(m-1)}) 
+ \frac{\lambda_1}{\tau} \sum_{l=1}^p 
\mathrm{I}_{\left\{|\hat{\beta}_{l}^{(m-1)}|\geq \tau \right\} } \cdot |\beta_l| 
+ \frac{\lambda_2}{\tau} \sum_{l<l':(l, l')\in \mathcal{E}} \mathrm{I}_{ \left\{ |\hat{\beta}_{l}^{(m-1)} - \hat{\beta}_{l'}^{(m-1)}|\geq \tau \right\} } 
\cdot |\beta_l - \beta_{l'}| \\
& \quad \qquad \qquad + \lambda_3 \sum_{l=1}^p 
\mathrm{I}_{\left\{ \hat{\beta}_{l}^{(m-1)} \geq 0 \right\} } \cdot \beta_l^2 \ .
\end{align*}

For the $m$-th approximation, an upper convex approximating function to $S(\bmbeta)$ can be obtained by $S^{(m)}(\bmbeta) = S_1(\bmbeta) - S_2^{(m)}(\bmbeta)$, which formulates the following subproblem.
\begin{equation*}
\min_{\bmbeta} \sum_{i=1}^n (y_i - \bx_i^T\bmbeta)^2 
+ \lambda \sum_{l=1}^{p} \beta_l^2
+ \frac{\lambda_1}{\tau} \sum_{l \in \mathcal{F}^{(m-1)}} |\beta_{l}| 
+ \frac{\lambda_2}{\tau} \sum_{l<l': (l,l')\in \mathcal{E}^{(m-1)} } |\beta_l - \beta_{l'}| 
+ \lambda_3 \sum_{l \in \mathcal{N}^{(m-1)} } \beta_l^2
\ ,
\end{equation*}
where 
\begin{equation}\label{eq:EFN-update}
\begin{split}
\mathcal{F}^{(m-1)} &= \left\{l: |\hat{\beta}_l^{(m-1)}|<\tau \right\}\ , \\
\mathcal{E}^{(m-1)} &= \left\{(l,l')\in \mathcal{E}, \ 
|\hat{\beta}_l^{(m-1)} - \hat{\beta}_{l'}^{(m-1)}|<\tau \right\}\ , \\
\mathcal{N}^{(m-1)} &= \left\{l: \hat{\beta}_l^{(m-1)} < 0 \right\}\ . 
\end{split}
\end{equation}

Denote $\beta_{ll'} = \beta_l - \beta_{l'}$ and define $\bmxi = (\beta_1, \cdots, \beta_p, \beta_{12}, \cdots, \beta_{1p}, \cdots, \beta_{(p-1)p})$. The $m$-th subproblem can be reformulated as an equality-constrained convex optimization problem, 
\begin{equation*}
\min_{\bmxi} \sum_{i=1}^n (y_i - \bx_i^T\bmbeta)^2 
+ \lambda \sum_{l=1}^{p} \beta_l^2 
+ \frac{\lambda_1}{\tau} \sum_{l \in \mathcal{F}^{(m-1)}} |\beta_l| 
+ \frac{\lambda_2}{\tau} \sum_{l<l': (l,l')\in \mathcal{E}^{(m-1)}} | \beta_{ll'}| 
+ \lambda_3 \sum_{l \in \mathcal{N}^{(m-1)} } \beta_l^2 \ ,
\end{equation*}
subject to $ {\beta_{ll'} = \beta_l - \beta_{l'}, \quad  \forall l<l':(l,l')\in \mathcal{E}^{(m-1)} } $.

The only difference for nnFGSPCA is the updating rule of 
$\hat{\beta}_{l}^{(m,k)}$, since $p_3(\bmbeta)$ does not involve other variables but only $\beta_l$. 
In particular, when updating by 
\begin{equation*}
\hat{\beta}_{l}^{(m,k)} = \alpha^{-1} \gamma .
\end{equation*}
The new $\alpha$ is formulated as follows,
\begin{equation}
\alpha = 2\lambda 
+ 2\lambda_3 \mathrm{I}_{\left \{\hat{\beta}_l^{(m-1)}<0 \right\}}
+ 2\sum_{i=1}^n x_{il}^2
+ \nu^{(k)} \left\vert l':(l,l') \in \mathcal{E}^{(m-1)} \right\vert ,
\end{equation}
where $\alpha$ is different compared to the solution of FGSPCA and $\gamma$ stays the same.




\end{document}